\documentclass[prd, aps, nofootinbib, preprint, letterpaper]{revtex4-2}

\usepackage{mathtools}
\usepackage{color}
\definecolor{darkred}{RGB}{196,0,0}

\usepackage{amsmath, amssymb, amsfonts}
\usepackage[hidelinks]{hyperref}
\usepackage{ulem}
\usepackage{setspace}
\usepackage{extarrows}
    
\newcommand{\beq}{\begin{equation}}
\newcommand{\eeq}{\end{equation}}
\newcommand{\bqa}{\begin{eqnarray}}
\newcommand{\eqa}{\end{eqnarray}}

\newcommand{\la}{\label}

\newcommand{\wh}{{\hat\omega}}

\begin{document}

\title {Suppression of the jet quenching parameter near the critical temperature }

\author{Haibo Ren$^{a,b}$, Qianqian Du$^{a,b}$, and Yun Guo$^{a,b,*}$}
\affiliation{
$^a$ Guangxi Key Laboratory of Nuclear Physics and Technology, Guangxi Normal University, Guilin 541004, China\\
$^b$ Department of Physics, Guangxi Normal University, Guilin, 541004, China
}

\renewcommand{\thefootnote}{\fnsymbol{footnote}}
\footnotetext[1]{Contact author: yunguo@mailbox.gxnu.edu.cn}
\renewcommand{\thefootnote}{\arabic{footnote}}

\begin{abstract}
In this work, we study the jet quenching parameter ${\hat q}$ by using a background field effective theory. Particular attention is paid to its behavior near the critical temperature where nonperturbative effects induced by the deconfining phase transition are taken into account through a self-consistently introduced background field ${\cal Q}$. 
We adopt a theoretical approach in which the interaction rate between the energetic jet and medium partons is computed diagrammatically and the hard-thermal-loop resummed propagator is used to regulate the infrared divergence. 
In the presence of a background field, its influence on the jet quenching parameter manifests in two aspects. One is the modification on the screening mass in the resummed propagator, which leads to an enhanced ${\hat q}$. The other corresponds to the ${\cal Q}$-modified parton distribution function which is dominant and leads to a suppression of ${\hat q}$. Decreasing the temperature $T$, our result shows a nonmonotonic $T$ dependence of the dimensionless ${\hat q}/T^3$. In the high temperature region, ${\hat q}/T^3$ shows an increase with decreasing $T$ due to the running coupling effect. Near the critical temperature, the background field plays a significant role and a dramatic suppression of ${\hat q}/T^3$ is found which qualitatively agrees with the lattice simulation. In addition, the background field modification on the jet quenching parameter 
which is characterized by the ${\hat q}$ ratio can be simply parametrized by a polynomial expression depending only on the background field. This expression is expected to be useful for phenomenological applications in jet physics.
\end{abstract}

\maketitle
\newpage

\section{Introduction}\la{intro}

The hot and dense quark-gluon plasma (QGP) produced in the relativistic heavy-ion collider (RHIC) and the Large Hadron Collider (LHC) is a novel state of matter whose properties cannot be directly observed and must be probed indirectly. Among various probes, the energy loss of a highly energetic parton as it traverses the strongly interacting medium, known as jet quenching, serves as an important tool for studying the transport properties of the QGP~\cite{dEnterria:2009xfs,Wiedemann:2009sh,Majumder:2010qh,Mehtar-Tani:2013pia,Qin:2015srf}. In ultrarelativistic heavy-ion experiments, the jet quenching phenomenon is observed as the suppression of high-transverse-momentum hadrons and the modification of di-hadron momentum balance, both of which are governed by the so-called jet quenching parameter, often denoted as ${\hat q}$. It measures the average squared transverse momentum transfer per unit path length to a fast-moving parton due to elastic collisions with the constituent quarks and gluons in the plasma.

The fundamental importance of the jet quenching parameter stems from the multiple roles it plays within the jet quenching framework. First of all, $\hat{q}$ is a key element that goes into the calculation of the jet energy loss. The radiative energy loss ${\mathrm{\Delta}E}_{\rm rad}$ depends directly on this parameter. For instance, when the medium length $L$ is much larger than the mean free path, it was found that ${\mathrm{\Delta}E}_{\rm rad} \approx \alpha_{s}\hat{q}L^{2}$ for soft gluon emissions~\cite{Baier:1996kr,Baier:1996sk}. In addition, $\hat{q}$ is related to the elastic scattering cross section, also known as the collision kernel, allowing calculation of the collisional energy loss which is more relevant to a heavy flavor jet. Second, there is an intrinsic connection between $\hat{q}$ and the shear viscosity $\eta$. In the weak coupling limit, $ T^{3}/{\hat{q}} \approx \eta/s $ where $T$ is the temperature of the plasma and $s$ is the entropy density. A derivation of this relation can be regarded as a quantitative measure of the strongly coupled nature of the QGP~\cite{Majumder:2007zh}. Furthermore, $\hat{q}$ is also an important parameter to determine whether the color coherence is maintained among partons within the jet, thereby influencing their radiation patterns and energy loss mechanisms~\cite{Mehtar-Tani:2011hma}. 

In the hot and dense QGP, the jet quenching parameter incorporates both the thermodynamic and dynamical properties of the medium, the extraction of $\hat{q}$ primarily relies on the experimental measurements of the nuclear modification factors associated with the above mentioned jet quenching phenomenon~\cite{JET:2013cls,JETSCAPE:2021ehl}. For example, with the jet quenching model based on the higher-twist parton energy loss formalism, a direct comparison to the measured single inclusive hadron nuclear modification factor $R_{AA}$ at RHIC energies leads to ${\hat q}\approx 1.2\, {\rm GeV}^2/{\rm fm}$ for a quark jet~\cite{JET:2013cls}. Theoretically, the jet quenching parameter has been extensively studied before. In the weak-coupling perturbative QCD, $\hat{q}$ was calculated at the leading-order in Refs.~\cite{Arnold:2008vd,Arnold:2008zu,Aurenche:2002pd} where the corresponding results applied to different regions of the transferred transverse momentum $q_\perp$. A general result that is valid for arbitrary $q_\perp$ has been obtained in Ref.~\cite{Boguslavski:2024jwr} where an effective propagator is adopted to deal with the hard and soft momentum transfer in a unified framework. The next-to-leading order result for $\hat{q}$ was obtained in Ref.~\cite{Caron-Huot:2008zna} which indicates a non-negligible higher order correction at a realistic coupling in QCD. Within the AdS/CFT framework, calculations of the jet quenching parameter have been performed at leading order in the inverse coupling for the strongly coupled ${\cal N}=4$ supersymmetric Yang-Mills plasma~\cite{Liu:2006he,DEramo:2010wup,Liu:2006ug}. Finite ’t Hooft coupling corrections which are found to reduce the leading order result have been discussed in Refs.~\cite{Armesto:2006zv,Zhang:2012jd}. Furthermore, lattice simulation on ${\hat q}$ has also been carried out in Ref.~\cite{Kumar:2020wvb} and the obtained ${\hat q}/T^3$ drops very quickly as the temperature $T$ approaches to the critical temperature. The simulation results also show some discrepancy as compared to the calculations using perturbation theory. In addition, there are many other studies on the jet quenching parameter which either used different theoretical approaches or considered various medium modifications. Some examples can be found in Refs.~\cite{Wang:2000uj,Antonov:2007sh,Baier:2008js,Ghiglieri:2015ala,Ghiglieri:2018ltw,Schlichting:2020lef,Moore:2021jwe,Grishmanovskii:2022tpb,Song:2022wil,Mehtar-Tani:2022zwf}.

The jet quenching parameter is not a static quantity as it depends on factors like jet energy, medium temperature, etc. Near the deconfining phase transition from the QGP to hadronic matter, the above mentioned dramatic decline in ${\hat q}/T^3$ from lattice QCD indicates that ${\hat q}$ is a nonperturbative parameter whose behavior cannot be simply obtained by using the traditional or the hard-thermal-loop (HTL) improved perturbation theory. Although the latter can provide reliable predictions on the thermodynamics of the plasma down to several times the critical temperature $T_d$~\cite{Andersen:2011sf,Haque:2014rua}, it fails to describe the rapid change of the thermodynamic quantities in the phase transition region. Studying the physics in a temperature region from $T_d$ to about $2\sim 3T_d$ which is termed as semi-QGP~\cite{Hidaka:2008dr} becomes very challenging. On the other hand, for SU$(N)$ gauge theories, the matrix models for deconfinement~\cite{Dumitru:2010mj,Dumitru:2012fw,Guo:2014zra} can well reproduce the lattice simulations on the thermodynamics in the entire semi-QGP region. In addition, these models also predict a deconfining phase transition through a $T$-dependent background field ${\cal Q}$ which is introduced for the timelike component of the gauge field $A_\mu$. Therefore, the matrix models for deconfinement are also known as the background field effective theory. As the order parameter for the deconfining phase transition, the Polyakov loop is directly related to the background field and the partial deconfined nature of the semi-QGP is thus characterized by an order parameter with values being nonzero but less than unity. 
Within the background field effective theory, the HTL resummed gluon propagator has been derived which leads to the ${\cal Q}$-modified screening masses~\cite{Guo:2020jvc}. Then it has been adopted to study the collisional energy loss of a heavy-quark~\cite{Lin:2013efa,Du:2024riq}, the in-medium properties of the quarkonium states~\cite{Liu:2024fki} as well as the shear and bulk viscosity of the QGP~\cite{Debnath:2025qhd}. In this work, we will consider the jet quenching parameter with an emphasis on its physical properties in the phase transition region where the effective theory is expected to provide a reliable result.

The rest of the paper is organized as follows. In Sec.~\ref{defel}, we present a general approach to computing the jet quenching parameter which takes into account all the momentum transfers in a unified framework and can be straightforwardly generalized to the effective theory of a semi-QGP. To assess the dependence of our results on the calculational scheme, we also make a systematic comparison of the jet quenching parameter obtained in different theoretical approaches. After reviewing some basics of the background field effective theory in Sec.~\ref{effth}, we provide a detailed derivation of ${\hat q}$ in the presence of a background field in Sec.~\ref{inrabf} with special attention paid to the nontrivial color structures of the corresponding results. The numerical results of ${\hat q}$ are presented in Sec.~\ref{nre} where we analyze how a non-zero background modifies the jet quenching parameter, especially in the nonperturbative semi-QGP region and also compare our results with the previously obtained ones as well as the lattice data. Finally, a summary can be found in Sec.~\ref{summary}.

\section{The theoretical approaches to computing the jet quenching parameter}\la{defel}

When a high-energy jet propagates through a plasma consisting of quarks and gluons in thermal equilibrium, multiple scattering with thermal partons leads to the transverse momentum broadening which is quantified by the jet quenching parameter $\hat{q}$. It presents the rate of squared transverse momentum transferred to the energetic jet and is defined by 
\beq
    \hat{q}(\Lambda_\perp)=\int_{q_\perp < \Lambda_\perp}{d^2{\bf q}_\perp} \, \frac{d\Gamma_{el}}{d^2 {\bf q}_\perp} \, q_\perp^2\, ,
\eeq
where $\Gamma_{el}$ is the interaction rate of elastic scatterings and $q_\perp$ is the transferred transverse momentum in each single scattering. In addition, a cutoff scale $\Lambda_\perp$ should be introduced in the above integral to regulate the ultraviolet divergence. 

The key quantity to determine the transverse momentum broadening is the differential cross section that has been extensively studied before. Notice that both the differential cross section and $\hat{q}$ suffer from an infrared divergence when the soft limit of the momentum transfer is considered. Consequently, the differential cross section is usually determined for large and small $q_\perp$ separately. For $q_\perp \gtrsim T$, the computation can be carried out in the diagrammatic approach by using the bare propagator, while for soft processes $q_\perp \ll T$, the HTL resummed propagator was used to eliminate the infrared divergence. For a high-energy massless quark jet propagating through a SU(3) gluonic plasma, the corresponding results have the following limiting forms 
\begin{align}\label{inter}
    \frac{d\Gamma_{\text{el}}}{d^2 q_\perp} \simeq \frac{1}{3\pi^2} \times
    \begin{cases}
    \dfrac{g^2 T m_D^2}{q_\perp^2 (q_\perp^2 + m_D^2)}\quad&{\rm for}\quad q_\perp \ll T\,, \\[2ex]
    \dfrac{6I(q_\perp/T)}{q_\perp^4}\dfrac{g^4 T^3}{\pi^2}\quad&{\rm for} \quad q_\perp \gtrsim  T\,,
    \end{cases}
\end{align}
where $m_D=g T$ is the perturbative Debye mass and 
\beq
    I(q_\perp/T) =\frac{2\pi^3}{T^3}\int \frac{dq_z}{2\pi} \int \frac{d^3k}{(2\pi)^3} \frac{|k-k_z|^2}{k(k-q_z)} f(k) [1 + f(k-q_z)] \delta(q_z + |\mathbf{k} - \mathbf{q}| - k)\, ,
\eeq
with the $z$ axis being the direction of the quark jet. In the case of the small cutoff where $\Lambda_\perp\ll T$, we can get the following result for ${\hat q}$ by using the first case of Eq.~(\ref{inter}):
\beq\label{qs}
     \hat{q}_{\rm soft}=\frac{g^2 C_R T m_D^2}{4\pi} \ln\left(1 + \frac{\Lambda_\perp^2}{m_D^2}\right)\, .
\eeq
When using the second case in Eq.~(\ref{inter}) to compute ${\hat q}$, one should introduce a lower cutoff $\lambda_\perp$ with $\lambda_\perp \gg m_D$ to regulate the infrared divergence. In general, the corresponding ${\hat q}$ needs to be evaluated numerically. However, under certain approximation,\footnote{As done in Ref.~\cite{Arnold:2008vd}, when $q_\perp < \lambda_\perp$, the approximation $I(q_\perp/T)\approx I(0)$ was used to get $\hat{q}_{\rm hard2}$ in Eq.~(\ref{qh}).} an analytical expression can be obtained as follows~\cite{Arnold:2008vd}:
\bqa\label{qh}
    \hat{q}_{\rm hard1}&=&\frac{4 g^4T^3}{\pi^3}\zeta(3) \ln \left(\frac{\Lambda_\perp}{\lambda_\perp}\right)\,,\nonumber\\
    \hat{q}_{\rm hard2}&=&\frac{4 g^4T^3}{\pi^3}\left\{\frac{\zeta(2) - \zeta(3)}{2}\left[1 -2\gamma_E +2\ln 2- 2\ln\left( \frac{\lambda_\perp}{T} \right) \right] - \sigma\right\}\,. 
\eqa
In the above equations, $\hat{q}_{\rm hard}$ has been decomposed into two parts $\hat{q}_{\rm hard1}$ and $\hat{q}_{\rm hard2}$ which correspond to the contributions linear and quadratic in the distribution function, respectively. In addition, $\zeta(x)$ is the Riemann-zeta function and $\gamma_E$ is the Euler-Mascheroni constant and $\sigma=\sum_{k=1}^{\infty}\ln [(k-1)!]/k^3\approx 0.386$. It is interesting to point out that in the weak-coupling limit where $m_D \ll \lambda_\perp \ll T$ can be satisfied, the $\lambda_\perp$ dependence is exactly canceled when summing up the soft and hard parts, leaving ${\hat q}\equiv\hat{q}_{\rm hard}+\hat{q}_{\rm soft}$ depends only on $\Lambda_\perp$. Notice that $\lambda_\perp$ should be understood as a separation scale now. To see the cancellation, we should replace $\Lambda_\perp$ with $\lambda_\perp$ in Eq.~(\ref{qs}) and drop the constant $1$ in the logarithm. 

In this work, we present a general approach to determining the jet quenching parameter $\hat{q}$ by taking into account all the momentum transfers in a unified framework. The key point is to use the HTL resummed gluon propagator to compute the interaction rate that ensures a finite result in the infrared limit without introducing an artificial cutoff to separate the hard and soft scatterings. This approach has been used in our recent work to calculate the collisional energy loss of a heavy quark where it has been shown that $|{\cal M}|^2$ is independent on the specific gauge one chooses~\cite{Cai:2025ntf}.

The interaction rate $\Gamma_{el}$ for the elastic scattering between the quark jet and thermal gluons is given by
\bqa\label{ir}
    \Gamma_{el}&=& \frac{1}{2E} \int \frac{d^3\mathbf{p}'}{(2\pi)^3 2E'} \int \frac{d^3\mathbf{k}}{(2\pi)^3 2k} n(k) \int \frac{d^3\mathbf{k}'}{(2\pi)^3 2k'} (1 + n(k')) \nonumber\\
    & \times& (2\pi)^4 \delta^4(P + K - P' - K')\left( \frac{1}{2} \sum_{\mathrm{spins}} \frac{1}{3} \sum_{\mathrm{color}} |{\cal M}|^2 \right).
\eqa
In the above equation, $P = (E, \mathbf{p})$ and $P^\prime = (E^\prime, \mathbf{p}^\prime)$ denote the four-momenta of the incoming and outgoing high-energy quark jet. For the thermal gluons that scatter off the quark jet, the incoming and outgoing momenta are given by $K = (k, \mathbf{k})$ and $K^\prime = (k^\prime, \mathbf{k}^\prime)$, respectively. $n(k)=(e^{k/T}-1)^{-1}$ is the Bose-Einstein distribution function of the medium partons and the phase space is also weighted by the final-state Bose enhancement factor $1 + n(k')$. Based on the tree-level Feynman diagrams for the quark-gluon scattering, we compute $|{\cal M}|^2$ in the above equation with the HTL resummed gluon propagator. As the result has no gauge dependence, we can only consider the gauge independent part of the HTL resummed gluon propagator $D^{\mu\nu}(Q)$. Explicitly, we have 
\beq\label{repro}
 i D^{\mu\nu}(Q) = {\Delta_T}(Q) g^{\mu\nu} +\left[( \hat{\omega}^2-1) {\Delta_T}(Q)- {\Delta_L}(Q)\right] M^\mu M^\nu \, ,
\eeq
where the momentum of the exchanged gluon is denoted by $Q=K^\prime-K=(\omega, {\bf q})$ and $M^\mu = (1, 0, 0, 0)$ is the four-velocity of the thermal rest frame. In addition,
\beq\label{dtl}
 {\Delta_T}(Q) = \frac{1}{Q^2 - \Pi_T(\hat{\omega})}\, ,\quad {\Delta_L}(Q)= \frac{1}{q^2 - \Pi_L(\hat{\omega})}\, ,
\eeq
with the dimensionless variable $\hat{\omega} \equiv \omega/q$. The transverse and longitudinal HTL gluon self-energies take the following forms
\bqa\label{pitl}
\Pi_T(\hat{\omega}) &=& \frac{m_D^2}{2} \hat{\omega}^2 \left[1 - \frac{\hat{\omega}^2 - 1}{2 \hat{\omega}} \ln \left(\frac{\hat{\omega} + 1 +i \epsilon}{\hat{\omega} - 1 +i \epsilon}\right)\right]\, , \nonumber \\
\Pi_L(\hat{\omega}) &=& m_D^2 \left[-1 + \frac{\hat{\omega}}{2} \ln \left(\frac{\hat{\omega} + 1+i \epsilon}{\hat{\omega} - 1+i \epsilon}\right)\right]\, .
\eqa

In general, the outcome takes a rather complicated form. By assuming the energy of the quark jet being much larger than all other momentum scales of the plasma, contribution from $t$-channel gluon exchange becomes dominant. The resulting squared matrix element in the high energy limit is actually the same as that obtained in Ref.~\cite{Cai:2025ntf} where further condition ${\bf p}=E {\hat {\bf v}}$ can be imposed for massless quark jet.\footnote{The simplified $|{\cal M}|^2$ applies as long as the energy and momentum of the incident parton are much larger than the temperature, $E, p\gg T$ and the momenta of the medium partons are on the order of temperature, $k,k^\prime \sim T$.} Explicitly, we have 
\bqa\label{mqg}
\frac{1}{6}\sum|{\cal M}|^2&=&128 g^4 E^{2}\Big\{ |\Delta_L(Q)|^2  \left[\mathbf{k} \cdot \mathbf{k'}  -(k - k')^{2}/4\right ]+ {\rm Re} \big(\Delta_L(Q) \Delta_T(Q)^*\big)\nonumber \\
&\times& \big[k\big({\hat {\bf v}} \cdot{\bf k}^\prime - ({\hat {\bf v}} \cdot {\hat {\bf q}})( {\bf k}^\prime \cdot {\hat {\bf q}}) \big)+k^\prime \big({\hat {\bf v}} \cdot{\bf k} - ({\hat {\bf v}} \cdot {\hat {\bf q}})( {\bf k}\cdot {\hat {\bf q}}) \big)\big]\nonumber \\
&+& |\Delta_T(Q)|^2\big[\big({\hat {\bf v}} \cdot{\bf k} - ({\hat {\bf v}} \cdot {\hat {\bf q}})( {\bf k} \cdot {\hat {\bf q}})\big)\big({\hat {\bf v}} \cdot{\bf k}^\prime - ({\hat {\bf v}} \cdot {\hat {\bf q}})( {\bf k}^\prime \cdot {\hat {\bf q}})\big)
\nonumber \\
&+&(k k^\prime-{\bf k} \cdot{\bf k}^\prime)\big(1 - ({\hat {\bf v}} \cdot {\hat {\bf q}})^2 \big)\big]\Big\}\,, 
\eqa
where ${\hat {\bf q}}= {\bf q}/q$.

For an isotropic plasma, the jet quenching parameter has no dependence on the direction of the momentum carried by the quark jet. Therefore, following the same strategy used in computing the collisional energy loss of a heavy quark in Refs.~\cite{Cai:2025ntf,Guo:2024mgh},  we can perform an average over all the possible directions of the unit vector ${\hat {\bf v}}$ which significantly simplifies the expression of ${\hat {q}}$. Consequently, the jet quenching parameter can be directly obtained based on the result of the collisional energy loss with a minor modification on the integrand. It involves replacing the factor $\omega$ representing the energy loss in each individual scattering with $q_\perp^2$ and setting the incident velocity $v=1$. Due to the replacement, a contribution that is quadratic in the distribution function survives now. As a result, we have to keep the Bose enhancement factor in the calculation of ${\hat q}$. Thus, we arrive at the following result
\bqa
    \hat{q}_{\rm I}&=& g^4 \int\frac{d^3{\bf k}}{(2\pi)^3k} \int\frac{d^3{\bf k}^\prime}{(2\pi)^3 (k+\omega)} (\pi q) (1-{\hat \omega}^2)\,n(k)\left(1 + n(k+\omega)\right) \theta(q^2-\omega^2) \nonumber\\
    &\times& \Big[|\Delta_T(Q)|^2(1-{\hat \omega}^2)^2\left((\omega+2k)^2+3q^2\right)
    +2|\Delta_L(Q)|^2\left((\omega+2k)^2-2q^2\right)\Big]\,.
\eqa
Here, we use $\hat{q}_{\rm I}$ to denote the jet quenching parameter obtained by fully reevaluate the squared matrix element with the HTL resummed gluon propagator. There is a simplified way to compute $|{\cal M}|^2$ which we will mention later and the corresponding result will be denoted as ${\hat{q}}_{\rm II}$ for the purpose of distinction. In the above equation, we have rewritten $q_\perp^2=q^2-\omega^2$ due to the energy conservation $\delta(\omega-{\hat{\bf v}}\cdot{\bf q})$. The integrand now depends on $k$, $q$ and $\omega$, and we should change the integral variables accordingly. Notice that in the high energy limit, ultraviolet divergence appears and a cutoff $\Lambda_\perp$ for the transverse component of the transferred momentum needs to be introduced. As a result, we should make the following changes on the integral variables
\beq\label{changeva}
\frac{1}{2(2\pi)^2}\int \frac{d^3 \mathbf{k}}{k} \int \frac{d^3\mathbf{k}'}{k^\prime}  \to \int_{0}^{\infty} dk \int_{-k}^{\infty} d\omega \int_{|\omega|}^{\min(2k+\omega, \sqrt{\omega^2+\Lambda_{\perp}^2})}q\, dq 
\, ,
\eeq
where the integral intervals are determined by simultaneously considering the constraints $\theta(q^2-\omega^2)$, $q_\perp^2<\Lambda_\perp^2$ and $-1\le \cos \theta_{{\bf k} {\bf k}^\prime} \le 1$. Our final result for ${\hat q}$ is given by
\bqa\label{qhat}
    \hat{q}_{\rm I}&=&\frac{g^4}{(2\pi)^3} \int_{0}^{\infty} dk \int_{-k}^{\infty} d\omega \int_{|\omega|}^{\min(2k+\omega, \sqrt{\omega^2+\Lambda_{\perp}^2})} dq\,q^2  (1-{\hat \omega}^2) n(k)\left(1 + n(k+\omega)\right) \nonumber\\
    &\times& \Big[|\Delta_T(Q)|^2(1-{\hat \omega}^2)^2\left((\omega+2k)^2+3q^2\right)
    +2|\Delta_L(Q)|^2\left((\omega+2k)^2-2q^2\right)\Big]\,.
\eqa

At this point, we would like to mention that as a widely used technique, the HTL resummation provides a feasible way to regulate the infrared divergence by including screening effect in the bare propagator. Therefore, as suggested in Ref.~\cite{Arnold:2002zm}, our strategy is to directly reevaluate the tree-level Feynman diagrams with the resummed propagator $D^{\mu\nu}(Q)$. On the other hand, a simple version to do the regulation is also proposed in the same reference where only the infrared sensitive part in the squared matrix element gets modified by the medium effects. It amounts to using the squared matrix element for the $t$-channel gluon exchange between massless scalar quarks which, up to a trivial color factor and coupling constant, is given by $|D^{\mu\nu}(Q)(P+P^\prime)_\mu(K+K^\prime)_\nu|^2$. This formula has been adopted in Ref.~\cite{Boguslavski:2024jwr} to study the jet quenching parameter. We have checked that after taking the high-energy limit and performing the average over ${\hat{\bf v}}$, the jet quenching parameter in Ref.~\cite{Boguslavski:2024jwr} can be obtained by adding two terms $4q^2 |\Delta_L(Q)|^2-4q^2 (1-{\hat \omega}^2)^2|\Delta_T(Q)|^2$ into the square bracket of Eq.~(\ref{qhat}) and will be denoted as ${\hat q}_{\rm II}$. These two terms vanish in the hard limit where $|\Delta_T(Q)|^2\rightarrow 1/Q^4$ and $|\Delta_L(Q)|^2\rightarrow 1/q^4$, while in the soft limit $k\gg q$ and $k\gg \omega$, they become higher order contributions and can be dropped. Therefore, the two different methods to regulate the infrared divergence lead to the same limiting results of ${\hat q}$. In addition, as shown in Ref.~\cite{Boguslavski:2024jwr}, the squared matrix element for quark-gluon scattering differs from that for quark-quark scattering only by a multiplicative color factor. However, in our approach where $|{\cal M}|^2$ is fully evaluated by using the resummed gluon propagator, in general, there is no such a simple proportionality~\cite{Cai:2025ntf}.

For hard momentum transfer $q_\perp \gtrsim T$, self-energy corrections to bare gluon propagator become negligible in Eq.~(\ref{qhat}) and it is obvious that our result of ${\hat q}$ can reproduce Eq.~(\ref{qh}) which is obtained by using the bare propagator. For soft momentum transfer $q_\perp \ll T$, our result coincides with that in Ref.~\cite{Boguslavski:2024jwr}, and thus the correct soft limit as given in Eq.~(\ref{qs}) can be reproduced based on Eq.~(\ref{qhat}). We also point out that to reproduce the soft limit for ${\hat q}$, one should assume $k \gg \omega$ so that the integral over $k$ can be analytically carried out. In addition, the integration variables $q$ and $\omega$ should be changed\footnote{After changing the integral variables, the lower bound of $k$ is on the order of $q_\perp$. For soft processes, we can set the lower bound to be zero~\cite{Boguslavski:2024jwr} in order to do the integral over $k$.} into ${\bf q}_\perp$ and ${\hat \omega}$, then one can make use of the sum rule~\cite{Aurenche:2002pd} to perform the integral over ${\hat \omega}$.

\begin{figure}[htbp]
\begin{center}        
\includegraphics[width=0.48\linewidth]{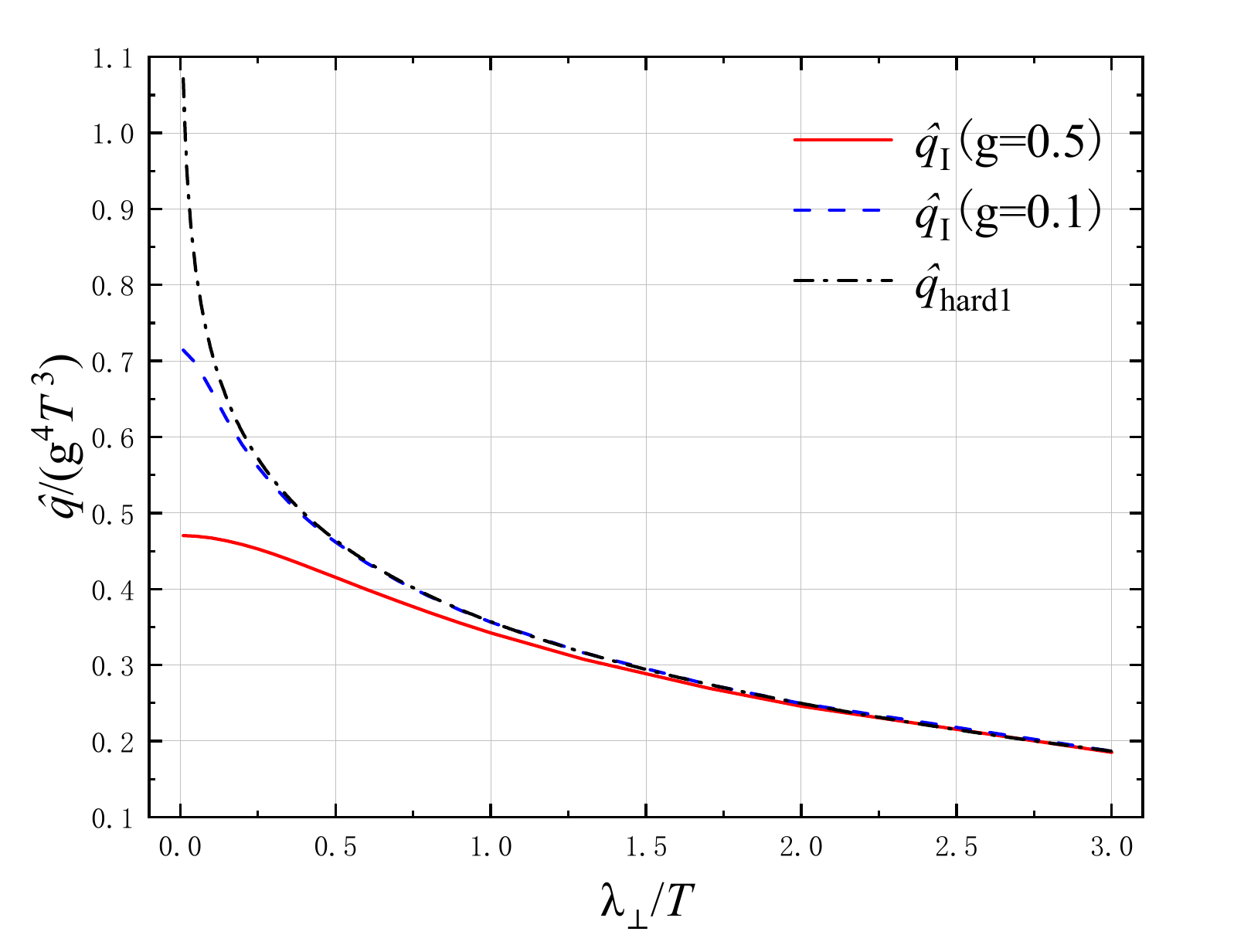}
\includegraphics[width=0.48\linewidth]{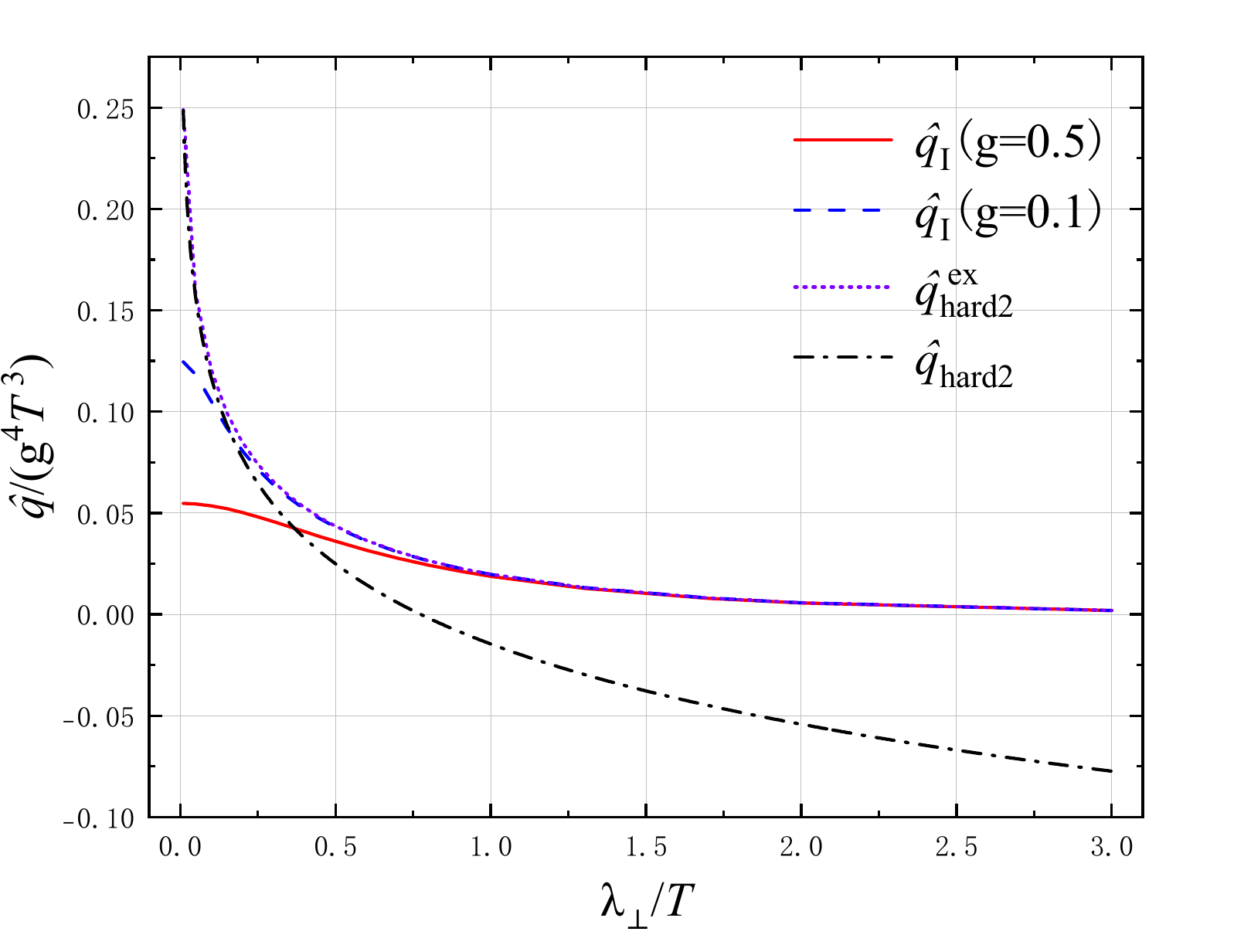}
\caption{The dimensionless $\hat{q}$ as a function of $\lambda_\perp/T$ at  $\Lambda_\perp/T=10$. The left plot shows the comparison for contributions linear in the distribution function obtained from Eq.~(\ref{qh}) and Eq.~(\ref{qhat}) at different gauge couplings. The right plot shows the corresponding comparison for contributions quadratic in the distribution function. ${\hat q}^{\rm ex}_{\rm hard2}$ presents the result obtained beyond the approximation used in ${\hat q}_{\rm hard2}$. See text for a more detailed discussion. }
\label{com1}
\end{center}
\end{figure}

In Fig.~\ref{com1}, analytical results for the hard contributions in Eq.~(\ref{qh}) are compared to the numerical evaluations of Eq.~(\ref{qhat}) with a lower bound $\lambda_\perp$ introduced for the transverse momentum $q_\perp$. We fix the upper bound $\Lambda_\perp/T=10$ and present how the dimensionless ${\hat q}$ depends on the lower cutoff for different values of the gauge coupling. In the left plot, we consider the hard contribution that is linear in the distribution function. It is found that for very small $\lambda_\perp$, discrepancies appear because the self-energy corrections start to play a role and using the resummed gluon propagator becomes necessary. As $\lambda_\perp$ increases, a good agreement appears which confirms that our general expression for the jet quenching parameter is reduced to Eq.~(\ref{qh}) when the lower bound of the momentum transfers gets large so that neglecting the self-energy correction is appropriate. Of course, this agreement depends on the coupling constant. For relatively large $g$, Debye screening effect is more pronounced, therefore, a larger cutoff $\lambda_\perp$ is needed to achieve a better agreement. 

The comparison for contributions quadratic in the distribution function is presented in the right plot in Fig.~\ref{com1}. Notice that Eq.~(\ref{qh}) is an approximated result, because it only works for $\lambda_\perp \ll T$ and gives rise to an unphysical negative contribution to ${\hat q}$ at large $\lambda_\perp$. Furthermore, to safely ignore the screening effect, $\lambda_\perp \gg g T$ must be satisfied. In this case, a good agreement between Eq.~(\ref{qhat}) and Eq.~(\ref{qh}) only exists in the weak coupling limit where $g T\ll \lambda_\perp \ll T$. The corresponding result beyond this approximation, denoted as ${\hat q}^{\rm ex}_{\rm hard2}$ is also plotted. It corresponds to evaluate the contribution quadratic in the distribution function based on Eq.~(\ref{inter}). When comparing with the result from Eq.~(\ref{qhat}) with the self-energy correction, very similar behaviors as found in the left plot can be seen.

\begin{figure}[htbp]
\begin{center}        
\includegraphics[width=0.48\linewidth]{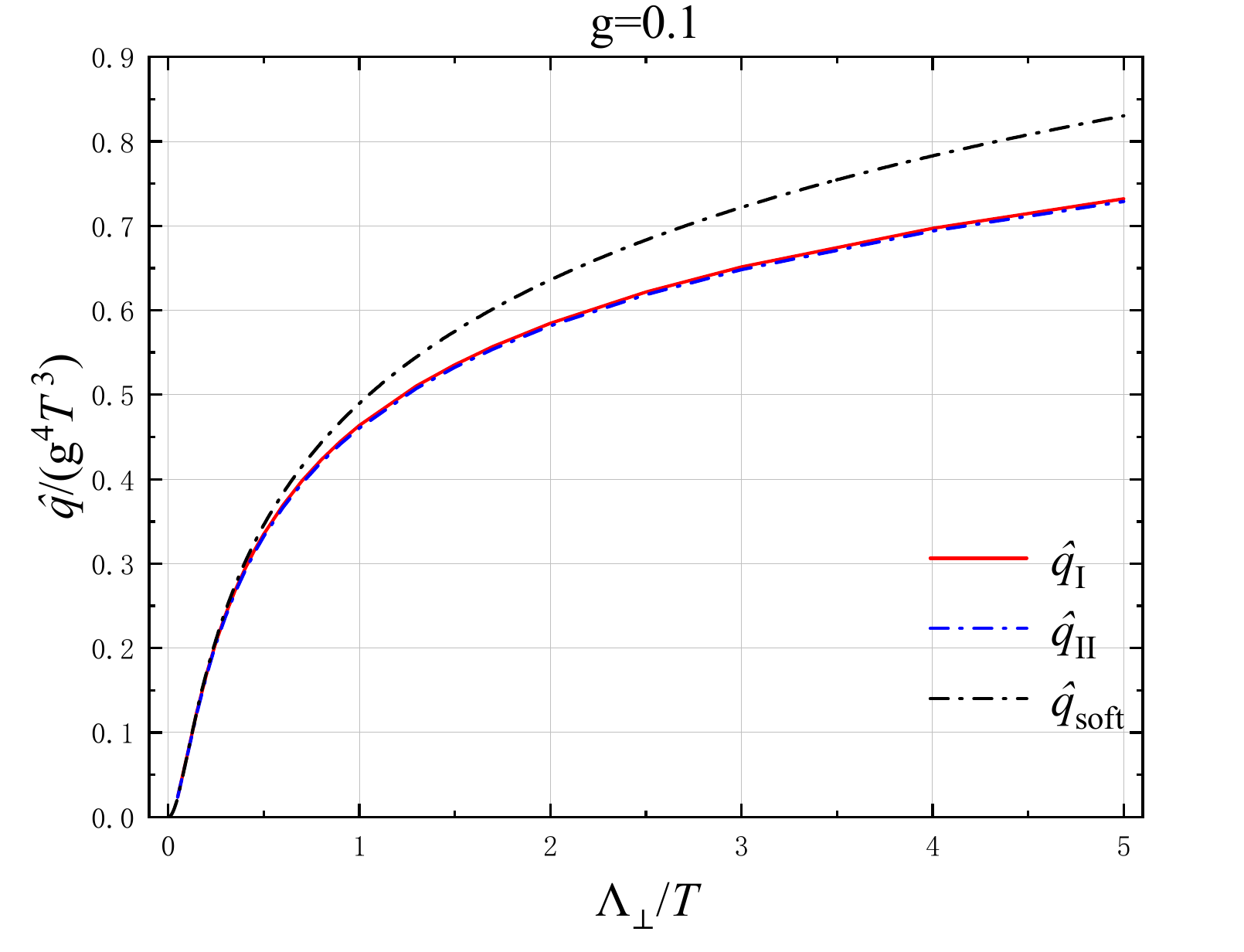}
\includegraphics[width=0.48\linewidth]{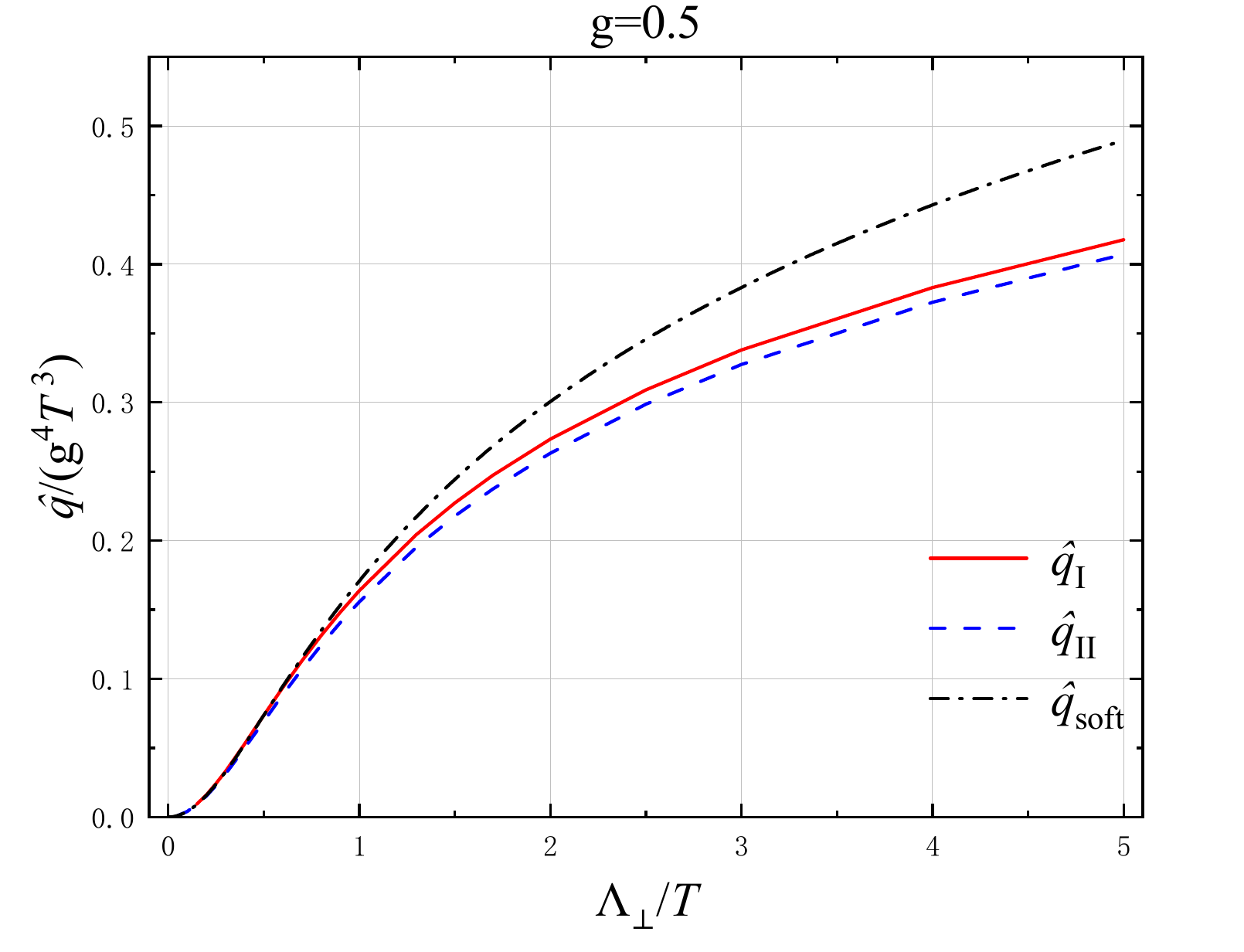}
\caption{The dimensionless $\hat{q}$ as a function of $\Lambda_\perp/T$ for two different gauge couplings. We plot the results from Eq.~(\ref{qhat}) and those obtained based on the simplified $|{\cal M}|^2$ as used in Ref.~\cite{Boguslavski:2024jwr}. In addition, the soft limit of ${\hat q}$ as given in Eq.~(\ref{qs}) is also shown for comparison.}
\label{com2}
\end{center}
\end{figure}

In Fig.~\ref{com2}, we show the dimensionless ${\hat q}$ as a function of $\Lambda_\perp/T$ for two different gauge couplings. Results from Eq.~(\ref{qhat}) and Ref.~\cite{Boguslavski:2024jwr} are presented for comparison. In addition, the corresponding soft limit of ${\hat q}$ as given in Eq.~(\ref{qs}) is also shown. As expected, consistent results are found for small cutoff $\Lambda_\perp$. However, a difference between our result and that in Ref.~\cite{Boguslavski:2024jwr} appears with increased $\Lambda_\perp$, which is more visible when the coupling constant gets large. Because the two results agree with each other in the hard and soft limit, it thus indicates that discrepancy at moderate $q_\perp$ still exists. On the other hand, the plots also verify that Eq.~(\ref{qs}) is a good approximation for $\Lambda_\perp< T$. With large cutoff, one should go beyond the soft limit in order to have a quantitatively reliable evaluation on the jet quenching parameter.

The above discussions suggest that 
for a realistic value of the gauge coupling in QCD, we can use either Eq.~(\ref{qhat}) or an alternative form as used in Ref.~\cite{Boguslavski:2024jwr} to get a quantitative estimate on the jet quenching parameter. However, it turns out to be that Eq.~(\ref{qhat}) can be generalized more straightforwardly to study the jet quenching parameter in a semi-QGP. Notice that unlike a trivial color factor in the above calculations, when considering a nonzero background field in semi-QGP, a much more complicated color structure appears and the resummed propagator is not simply diagonal in the color space. Given this fact, a full reevaluation of $|{\cal M}|^2$ is required to get the correct color structures. In addition, the validity of using resummed propagator to compute the squared matrix element has already been justified in Ref.~\cite{Cai:2025ntf} which shows that a gauge independent result can be obtained and unphysical gluon polarizations can be also eliminated in a self-consistent way by using the ghost fields.

\section{The jet quenching parameter in a semi-QGP}
\la{elinsqgp}

In this section, we study the jet quenching parameter in a semi-QGP with focus on the influence of the background field ${\cal Q}$ on this parameter. In Sec.~\ref{effth}, we briefly review the effective theory of a semi-QGP adopted in our calculation. More details can be found in Refs.~\cite{Hidaka:2020vna,Guo:2020jvc}. In Sec.~\ref{inrabf}, we derive an expression for ${\hat q}$ by using the ${\cal Q}$-modified HTL resummed gluon propagator where special attentions are paid to the highly nontrivial color structures. In Sec.~\ref{nre}, the numerical results of $\hat{q}$ in the presence of a background field are discussed.

\subsection{Effective theory of a semi-QGP}\label{effth}

For SU$(N)$ gauge theories, the nontrivial behavior of the Polyakov loop denoted as $\ell$ can be studied through a classical background field $A_0^{\rm cl}$ introduced for the gauge field. It is assumed to be constant in spacetime and given by a traceless matrix that is diagonal in color space, $(A_0^{\rm cl})_{ab}={\cal Q}^a \delta_{ab}/g$, with $\sum_{a=1}^{N}{\cal Q}^a=0$ for $a,b=1,\cdots, N$. In the presence of a background field, one can compute the effective potential, based on which ${\cal Q}$ is self-consistently determined from its equation of motion. We expect that a proper $T$ dependence of the background field will give a reasonable description of the Polyakov loop that characterizes the deconfining phase transition. 

However, within perturbation theory, the equation of motion predicts a vanishing ${\cal Q}$, therefore, the one-loop effective potential corresponds to the free energy of an ideal gas consisting of massless gluons and the system always stays in a completely deconfined phase where $\ell =1$~\cite{Dumitru:2012fw}. In fact, including the two-loop corrections~\cite{Dumitru:2013xna,Guo:2018scp}, we still have ${\cal Q}=0$ as the vacuum. To drive the system to confinement, we need to go beyond the perturbation framework and consider effective theories. Among different approaches to constructing a nonperturbative term for the effective potential, we consider a contribution which is generated by embedding a two-dimensional ghost field isotropically in four dimensions and imposing an upper limit $\sqrt{C}$ on the transverse momentum of
the embedded fields~\cite{Hidaka:2020vna}. Such a nonperturbative term favors a confined vacuum with $\ell=0$. Adding it to the one-loop effective potential, deconfining phase transition arises through a competition between the two different contributions making up the effective potential.

The background field determined from its equation of motion in the effective theory shows a $T$ dependence which qualitatively leads to the desired Polyakov loop. In the high-temperature limit, the vanishing ${\cal Q}$ indicates a completely deconfined phase with $\ell=1$ , while at low temperatures, ${\cal Q}$ is a constant for a given color number $N$ and $\ell=0$ suggests a confined vacuum. By requiring the phase transition happens at the critical temperature $T_d$, the parameter $C$ associated with the nonperturbative term in the effective theory can be fixed. To be specific, we consider SU(3) gauge theory and parametrize the background field as ${\vec {\sf q}}=({\sf q}, 0, - {\sf q})$ with ${\sf q}^a\equiv {\cal Q}^a/(2\pi T)$. In this case, the explicit $T$-dependent form of the background field in the deconfined phase is given by~\cite{Guo:2020jvc}
\beq\label{bf}
{\sf q}=\frac{1}{36}\big(9-\sqrt{81-80 (T_d/T)^2}\big)\, ,
\eeq
where we used $\sqrt{C}/T_d =2\sqrt{10} \pi/9$.

Both the parton distribution functions and the resummed gluon propagator which are relevant to our study on the jet quenching parameter will be modified by the background field. It acts as an imaginary chemical potential in the distribution functions and the resulting Bose-Einstein distribution function now reads~\cite{Furuuchi:2005zp,Hidaka:2008dr}
\begin{equation}\la{bd}
n(k_0,{\sf q}^{ab})=\left\{
\begin{aligned}
&\frac{1}{e^{|k_0|/T- 2 \pi i {\sf q}^{ab}}-1}\equiv n_{+}^{ab}(k_0) \quad\quad {\rm for}\quad k_0>0\, ,\\
&\frac{1}{e^{|k_0|/T+2 \pi i {\sf q}^{ab}}-1}\equiv n_{-}^{ab}(k_0) \quad\quad {\rm for}\quad k_0<0 \, ,\\
\end{aligned}
\right.
\end{equation}
with $ {\sf q}^{ab}\equiv {\sf q}^{a}- {\sf q}^{b}$. On the other hand, the resummed gluon propagator can be derived from the Dyson-Schwinger equation with a ${\cal Q}$-modified gluon self-energy. We point out that in the presence of a background field, it is more convenient to perform the calculation in the double line basis~\cite{Hidaka:2009hs,Cvitanovic:1976am} where color indices in the fundamental representation are denoted by $a,b,\cdots = 1,2,\cdots, N$, while in the adjoint representation, they are denoted by a pair of fundamental indices, $ab$. There are $N^2$ generators of the fundamental representation which are overcomplete. Consequently, individual components (in color space) of the propagator cannot be uniquely determined. A previous study~\cite{Guo:2020jvc} has already shown that ambiguities only show up in the determination of the diagonal components of the resummed gluon propagator, while the off-diagonal components have unique forms as the following\footnote{This result is obtained in covariant gauge with gauge parameter $\xi=0$.}:
\beq
i \tilde{D}_{\mu\nu}^{ab,cd}(Q,{\vec {\sf{q}}}) \xlongequal[]{a\neq b} \delta^{ad}\delta^{bc}\left[{\tilde \Delta}_T^{(ab)}(Q,{\vec {\sf{q}}}) A_{\mu\nu}+{\tilde \Delta}_L^{(ab)}(Q,{\vec {\sf{q}}}) B_{\mu\nu}\right]\, ,
\label{off_diagonal_propagator}
\eeq
where 
\beq
{\tilde \Delta}_T^{(ab)}(Q,{\vec {\sf{q}}})=[Q^2-{\tilde \Pi}^{(ab)}_{T}(\wh,{\vec {\sf{q}}})]^{-1}\,,\quad{\tilde \Delta}_L^{(ab)}(Q,{\vec {\sf{q}}})=[q^2-{\tilde \Pi}^{(ab)}_{L}(\wh,{\vec {\sf{q}}})]^{-1}\,,
\eeq
and
\beq
   A_{\mu \nu}(Q) = g_{\mu \nu} - \frac{Q_{\mu} Q_{\nu}}{Q^{2}} - \frac{\tilde{M}_{\mu} \tilde{M}_{\nu}}{\tilde{M}^{2}}\, ,\quad B_{\mu \nu}(Q) = \frac{1}{\wh^2-1} \frac{\tilde{M}_{\mu} \tilde{M}_{\nu}}{\tilde{M}^{2}}\,,
\eeq
with $\tilde{M}_{\mu} = M_{\mu} - \omega Q_{\mu}/Q^2$ being the part orthogonal to $Q^{\mu}$. In addition, ${\tilde \Pi}^{(ab)}_{T}(\wh,{\vec {\sf{q}}})$ and ${\tilde \Pi}^{(ab)}_{L}(\wh,{\vec {\sf{q}}})$ are the transverse and longitudinal gluon self-energies in the background field effective theory which can be simply obtained from their counterparts at ${\cal Q}=0$ as given in Eq.~(\ref{pitl}) by replacing the perturbative Debye mass $m_D$ with its ${\cal Q}$-modified version denoted as ${\cal M}^{(ab)}_D$. This is because ${\cal Q}$ modifications on the gluon self-energy are entirely encoded in the ${\cal Q}$-dependent screening mass ${\cal M}^{(ab)}_D$ which is just a multiplication of the perturbative Debye mass and given by
\beq\label{offmod}
\frac{({\cal M}^{(ab)}_D)^2}{m_D^2}= \frac{3}{N}\sum_{e=1}^{N}\bigg(B_2(|{\sf q}^{ae}|)+B_2(|{\sf q}^{eb}|)\bigg)+\frac{3 C}{4\pi^2 T^2}\, ,
\eeq
with $a\neq b$. In the above equation, the second Bernoulli polynomial is given by $B_2(x)=x^2-x+1/6$. It is a periodic function of $x$, and the argument $x$ should be understood as $x - [x]$ with $[x]$ the largest integer less than $x$, which is nothing but the modulo function. For SU(3), there are six off-diagonal screening masses. Explicitly, we have
\beq
{\cal M}_D^{(23)}={\cal M}_D^{(32)}={\cal M}_D^{(12)}={\cal M}_D^{(21)}=m_D\sqrt{1+\beta+7 {\sf q}^2-5{\sf q}}\, ,\nonumber
\eeq
\beq\label{offmass}
{\cal M}_D^{(13)} = {\cal M}_D^{(31)}=m_D\sqrt{1+\beta+10{\sf q}^2-6{\sf q}}\,,
\eeq
with $\beta=3C/(4\pi^2T^2)=30T_d^2/(81T^2)$.

As we just mentioned above, due to the overcomplete generators adopted in our calculation, there exists ambiguities in determining the diagonal components of the resummed gluon propagator. In Ref.~\cite{Guo:2020jvc}, it was found that the color summation over ${\cal P}^{ab,cd}\tilde{D}_{\mu\nu}^{ab,cd}$ has a unique expression which has been computed for $N=2$ and $N=3$ explicitly. Here, the projection operator ${\cal P}^{ab,cd}$ is defined as ${\cal P}^{ab,cd}=\delta_{ad}\delta_{bc}-\delta_{ab}\delta_{cd}/N$. The corresponding result for SU(3) is
\beq
i\sum_{ab}{\cal P}^{aa,bb}\tilde{D}_{\mu\nu}^{aa,bb}(Q,{\vec {\sf{q}}}) = \sum_{i=1}^2\left[{\tilde \Delta}_T^{[i]}(Q,{\vec {\sf{q}}}) A_{\mu\nu}+{\tilde \Delta}_L^{[i]}(Q,{\vec {\sf{q}}})B_{\mu\nu}\right]\, ,
\label{diagonal_propagator}
\eeq
where
\beq\label{d12}
{\tilde \Delta}_T^{[i]}(Q,{\vec {\sf{q}}})=[Q^2-{\tilde \Pi}^{[i]}_{T}(\wh,{\vec {\sf{q}}})]^{-1}\,,\quad {\tilde \Delta}_L^{[i]}(Q,{\vec {\sf{q}}})=[q^2-{\tilde \Pi}^{[i]}_{L}(\wh,{\vec {\sf{q}}})]^{-1}\,.
\eeq
Similar to earlier results, the gluon self-energies ${\tilde \Pi}^{[1]}_{T/L}(\wh,{\vec {\sf{q}}})$ and ${\tilde \Pi}^{[2]}_{T/L}(\wh,{\vec {\sf{q}}})$ can be obtained from Eq.~(\ref{pitl}) by replacing $m_D$ with ${\cal M}^{[1]}_{D}$ and ${\cal M}^{[2]}_{D}$, respectively. Therefore, the ${\cal Q}$ modification enters in a similar manner as the off-diagonal components. Here, the ${\cal Q}$-modified screening masses for the two diagonal gluons can be written as
\beq\la{diamass}
{\cal M}^{[1]}_{D}= m_D\sqrt{1+\beta+ 6 {\sf q}^2-6{\sf q}}\,,\quad\quad
{\cal M}^{[2]}_{D}=m_D \sqrt{1+\beta+ 18 {\sf q}^2-10{\sf q}}\,.
\eeq
However, in the evaluation on the jet quenching parameter in semi-QGP, the color structure that associated with the diagonal components of the resummed gluon propagator is not the one as given in Eq.~(\ref{diagonal_propagator}). The completely new color structure that will appear in the following calculation makes this evaluation highly nontrivial. We have to work out this new color structure explicitly and show the corresponding result has a unique form although the individual component of $\tilde{D}_{\mu\nu}^{aa,bb}$ has not.

\subsection{The jet quenching parameter in the effective theory}\label{inrabf}

In this subsection, we derive the jet quenching parameter in a semi-QGP where a nonzero background field is present. The derivation will be carried out in the double line basis and the corresponding Feynman rules can be found in Refs.~\cite{Hidaka:2009hs,Wang:2022dcw}. Since we are interested in the nontrivial color structures, to make the expressions compact, we will simply drop the Lorentz indices as the Lorentz structures are not changed as compared to the case where ${\cal Q}$ vanishes. 

We start by considering the off-diagonal components of the ${\cal Q}$-modified resummed gluon propagator. Recall that the background field also modifies the parton distribution functions, so the weight factors $n(k)(1+n(k+\omega))$ will be included in the squared matrix element. Then, the following color sums need to be taken into account\footnote{In the following, the ${\cal Q}$-modified distribution function $n^{ab}(k)$ corresponds to $n^{ab}_{+}(k)$ in Eq.~(\ref{bd}).}
\bqa\label{m2off}
|{\cal M}|_{\rm off}^2&\sim& \sum_{\rm colors}^{e\neq f, g\neq h}  {\rm Tr}[t^{ef}t^{gh}] f^{ab,cd,ef}{\tilde D}^{fe,ef}f^{ba,dc,gh}({\tilde D}^{*})^{hg,gh}n^{ab}(k)(1 + n^{cd}(k+\omega))\nonumber\\
    &=&\frac{1}{4} \sum_{a}^{e \neq f} \left[ n^{ae}(k)(1 + n^{fa}(k+\omega)) + n^{ea}(k)(1 + n^{af}(k+\omega)) \right]{\tilde D}^{fe,ef}({\tilde D}^{*})^{ef,fe}\,,
\eqa
where we used ${\rm Tr}[t^{ab}t^{cd}]={\cal P}^{ab,cd}/2$ and the structure constant is defined as $f^{ab,cd,ef}=i (\delta^{ad}\delta^{cf}\delta^{eb}-\delta^{af}\delta^{cb}\delta^{ed})/\sqrt{2}$. As we can see, for a given pair of $ef$ with $e\neq f$, contributions to the squared matrix element from each of the $N^2-N$ off-diagonal gluons have the same form and the only difference lies in the ${\cal Q}$-modified screening masses and the sum of the distribution functions. The corresponding ${\hat q}$ can be derived after contracting the Lorentz indices which is another cumbersome calculation. Fortunately, this contraction can be done by following exactly the same procedure as the ${\cal Q}=0$ case. It is straightforward to get the contributions to ${\hat q}$ from exchange of the off-diagonal gluons,
\bqa\label{qhatoff}
    \hat{q}_{\rm off}&=&\frac{g^4}{6(4\pi)^3} \sum_{a}^{e\neq f}\int_{0}^{\infty} dk \int_{-k}^{\infty} d\omega \int_{|\omega|}^{\min(2k+\omega, \sqrt{\omega^2+\Lambda_{\perp}^2})} dq\,q^2  (1-{\hat \omega}^2)  \nonumber\\
    &\times&\Big[|{\tilde \Delta}_T^{(ef)}(Q,{\vec {\sf{q}}})|^2(1-{\hat \omega}^2)^2\left((\omega+2k)^2+3q^2\right)
    +2|{\tilde \Delta}_L^{(ef)}(Q,{\vec {\sf{q}}})|^2\left((\omega+2k)^2-2q^2\right)\Big]\nonumber\\
    &\times&\left[n^{ae}(k)(1 + n^{fa}(k+\omega)) + n^{ea}(k)(1 + n^{af}(k+\omega)) \right]\,.
\eqa
To get the above equation, we have taken into account the fact that the ghost contribution has the same color structure as the $t$-channel quark-gluon scattering. As compared to Eq.~(\ref{qhat}), we find that besides the simple replacement of the Debye mass with the ${\cal Q}$-modified ones, there is also a nontrivial change in the distribution functions.

The above result is valid for general SU$(N)$. For $N=3$, we have six off-diagonal gluons which acquire only two different screening masses according to Eq.~(\ref{offmass}). Therefore, after explicitly summing over the color indices, we can rewrite Eq.~(\ref{m2off}) as the following\footnote{For SU(3), there are only four different ${\cal Q}$-modified screening masses. We use $i=3$ and $i=4$ to denote the third and fourth ${\cal Q}$-modified screening masses associated with the off-diagonal gluons. The first two with $i=1$ and $i=2$ correspond to the screening masses associated with the two diagonal gluons, see Eq.~(\ref{diagonal_propagator}).}:
\bqa\label{m2offnew}
|{\cal M}|_{\rm off}^2\sim \frac{1}{2} \sum_{i=3}^4\Big[ |{\tilde \Delta}_T^{[i]}(Q,{\vec {\sf{q}}})|^2  A^{\rho \sigma} A^{\rho^\prime \sigma^\prime}+|{\tilde \Delta}_L^{[i]}(Q,{\vec {\sf{q}}})|^2  B^{\rho \sigma} B^{\rho^\prime \sigma^\prime}\Big] {\cal F}^{[i]}\,,
\eqa
where
\bqa
{\tilde \Delta}_T^{[3]}(Q,{\vec {\sf{q}}})&=&[Q^2-{\tilde \Pi}^{(12)}_{T}(\wh,{\vec {\sf{q}}})]^{-1}\, ,\quad {\tilde \Delta}_L^{[3]}(Q,{\vec {\sf{q}}})=[q^2-{\tilde \Pi}^{(12)}_{L}(\wh,{\vec {\sf{q}}})]^{-1}\,, \nonumber \\
{\tilde \Delta}_T^{[4]}(Q,{\vec {\sf{q}}})&=&[Q^2-{\tilde \Pi}^{(13)}_{T}(\wh,{\vec {\sf{q}}})]^{-1}\, ,\quad {\tilde \Delta}_L^{[4]}(Q,{\vec {\sf{q}}})=[q^2-{\tilde \Pi}^{(13)}_{L}(\wh,{\vec {\sf{q}}})]^{-1}\, ,
\eqa
and
\bqa
{\cal F}^{[3]}&=&\big(2n(k)+n^{13}(k)\big)\big(1+n^{21}(k+\omega)\big)+\big(2n(k)+n^{31}(k)\big)\big(1+ n^{12}(k+\omega)\big) \nonumber \\
&+&n^{12}(k)\big(3+2n(k+\omega)+ n^{31}(k+\omega)\big)+n^{21}(k)\big(3+2n(k+\omega)+ n^{13}(k+\omega)\big)\, ,\nonumber \\
{\cal F}^{[4]}&=&n^{12}(k)\big(1 + n^{12}(k+\omega)\big)+n^{21}(k)\big(1 + n^{21}(k+\omega)\big)+\big(n^{13}(k)+n^{31}(k)\big)\big(1+n(k+\omega)\big)\nonumber \\
&+&n(k)\big(2 + n^{13}(k+\omega)+ n^{31}(k+\omega))\big)\, .
\eqa
In Eq.~(\ref{m2offnew}), cross terms containing both transverse and longitudinal gluon self-energies are omitted because these terms have no contribution to ${\hat q}$ after averaging over the directions of the momentum carried by the quark jet. As a result, the jet quenching parameter takes the following form for SU(3), 
\bqa\label{qhatoffsu3}
    \hat{q}_{\rm off}&=&\frac{g^4}{3(4\pi)^3} \sum_{i=3}^{4}\int_{0}^{\infty} dk \int_{-k}^{\infty} d\omega \int_{|\omega|}^{\min(2k+\omega, \sqrt{\omega^2+\Lambda_{\perp}^2})} dq\,q^2  (1-{\hat \omega}^2) {\cal F}^{[i]}  \nonumber\\
    &\times&\Big[|{\tilde \Delta}_T^{[i]}(Q,{\vec {\sf{q}}})|^2(1-{\hat \omega}^2)^2\left((\omega+2k)^2+3q^2\right)
    +2|{\tilde \Delta}_L^{[i]}(Q,{\vec {\sf{q}}})|^2\left((\omega+2k)^2-2q^2\right)\Big]\,.
\eqa

Deriving the contributions from diagonal gluons to the jet quenching parameter is not that straightforward as the off-diagonal ones. We first show that the following new color structure showing up in the matrix element has an unique form,
\beq\label{im}
i{\cal M}\sim f^{ab,cd,ee}{\tilde D}^{ee,ff}t^{ff}\,,
\eeq
where sums over indices $e$ and $f$ will be performed in the case of SU(3). For example, with $f=1$, summing over $e$ yields 
\bqa
\sum_{e=1}^{3}f^{ab,cd,ee}{\tilde D}^{ee,11}t^{11}
&=&\big[f^{ab,cd,,11}({\tilde D}^{22,11}-{\tilde {\cal D}}^{22,11})+f^{ab,cd,,22}{\tilde D}^{22,11}+f^{ab,cd,,33}{\tilde D}^{33,11}\big]t^{11}\nonumber\\
&=&\big[-f^{ab,cd,11}{\tilde{\cal D}}^{22,11}+f^{ab,cd,33}({\tilde {\cal D}}^{33,11}-{\tilde {\cal D}}^{22,11})\big]t^{11}\nonumber\\
&=&\big[f^{ab,cd,22}{\tilde{\cal D}}^{22,11}+f^{ab,cd,33}{\tilde {\cal D}}^{33,11}\big]t^{11}\,,
\eqa
where we used $\sum_e f^{ab,cd,ee}=0$ and define ${\tilde {\cal D}}^{aa,bb}\equiv{\tilde D}^{aa,bb}-{\tilde D}^{bb,bb}$. Similarly, the results for $f=$2 and $f=3$ can be obtained. Therefore, summing over $e$ and $f$ in Eq.~(\ref{im}) can be rewritten as
\beq
i{\cal M}\sim \sum_{e\neq f}f^{ab,cd,ee}{\tilde {\cal D}}^{ee,ff}t^{ff}\, ,
\eeq
where $e=f$ is excluded in the color sums. One important consequence of this expression is that there is no ambiguity in determining the matrix element because unlike the individual component of ${\tilde D}^{aa,bb}$, the quantity ${\tilde {\cal D}}^{aa,bb}$ can be uniquely determined according to the discussions in our previous work~\cite{Guo:2020jvc}. For SU(3), the explicit results are given by
\bqa
i{\tilde {\cal D}}^{11,22}_{\mu\nu}(Q,{\vec {\sf{q}}})&=&i{\tilde {\cal D}}^{33,22}_{\mu\nu}(Q,{\vec {\sf{q}}})= {\tilde \Delta}_{T}^{[1]}(Q,{\vec {\sf{q}}})A_{\mu\nu}+{\tilde \Delta}_{L}^{[1]}(Q,{\vec {\sf{q}}})B_{\mu\nu} \, ,\nonumber \\
i{\tilde {\cal D}}^{33,11}_{\mu\nu}(Q,{\vec {\sf{q}}})&=&i{\tilde {\cal D}}^{11,33}_{\mu\nu}(Q,{\vec {\sf{q}}})= {\tilde \Delta}_{T}^{[2]}(Q,{\vec {\sf{q}}})A_{\mu\nu}+{\tilde \Delta}_{L}^{[2]}(Q,{\vec {\sf{q}}})B_{\mu\nu}\, ,\nonumber \\
{\tilde {\cal D}}^{22,11}_{\mu\nu}(Q,{\vec {\sf{q}}})&=&{\tilde {\cal D}}^{22,33}_{\mu\nu}(Q,{\vec {\sf{q}}})=\frac{1}{2}\big({\tilde {\cal D}}^{11,22}_{\mu\nu}(Q,{\vec {\sf{q}}})+{\tilde {\cal D}}^{33,11}_{\mu\nu}(Q,{\vec {\sf{q}}})\big)\, ,
\eqa
where ${\tilde \Delta}_{T/L}^{[i]}(Q,{\vec {\sf{q}}})$ has been defined in Eq.~(\ref{d12}). As we can see, the above expressions take very similar forms as the off-diagonal propagators in Eq.~(\ref{off_diagonal_propagator}).

Including the parton distribution functions, the color structure appearing in the squared matrix element can be expressed as
\bqa\label{m2dia}
    |{\cal M}|_{\rm dia}^2&\sim&\sum_{\rm colors}^{e\neq f,g\neq h}{\rm Tr}[t^{ff}t^{hh}]f^{ab,cd,ee}{\tilde {\cal D}}^{ee,ff}f^{ba,dc,gg}({\tilde {\cal D}}^{*})^{gg,hh}n^{ab}(k)(1+n^{cd}(k+\omega))\nonumber\\
    &=&\frac{1}{12}\sum_{a}^{e\neq f,g\neq h}{\tilde {\cal D}}^{ee,ff}({\tilde {\cal D}}^{*})^{gg,hh}(3\delta_{hf}-1)(\delta_{eg}-\delta_{ea})\big[n^{ag}(k)(1+n^{ga}(k+\omega))\nonumber\\
    &+&n^{ga}(k)(1+n^{ag}(k+\omega))\big]\,.
\eqa
Based on the above expression for the squared matrix element, it is far from obvious how one could recast it into a form similar as Eq.~(\ref{m2off}) so that contributions to the jet quenching parameter from each of the two diagonal gluons can be expressed by an analogous form as appears in the case of vanishing background field, see Eq.~(\ref{qhat}). Recall that for the off-diagonal gluons, we do have such an expression for ${\hat q}$ given by Eq.~(\ref{qhatoff}) where each of the six gluons is labeled by a pair of color index $ef$. Furthermore, if we consider ${\tilde {\cal D}}^{aa,bb}_{\mu\nu}$ instead of ${\tilde D}^{aa,bb}_{\mu\nu}$ as the propagators for diagonal gluons, naively we would expect that an unwanted structure $ \big({\tilde \Delta}_{T}^{[i]} \big) \big({\tilde \Delta}_{T}^{[j]}\big)^*$ with $i\neq j$ would show up which makes the final results rather complicated. While for off-diagonal components, only $i=j$ is allowed, see Eq.~(\ref{m2offnew}). Fortunately, using the explicit results of ${\tilde {\cal D}}^{ee,ff}$ for SU(3) and performing the sum over the color indices, a rather tedious but straightforward calculation shows the following expression for the squared matrix element,
\bqa
|{\cal M}|_{\rm dia}^2\sim \frac{1}{2} \sum_{i=1}^2\Big[ |{\tilde \Delta}_T^{[i]}(Q,{\vec {\sf{q}}})|^2  A^{\rho \sigma} A^{\rho^\prime \sigma^\prime}+|{\tilde \Delta}_L^{[i]}(Q,{\vec {\sf{q}}})|^2  B^{\rho \sigma} B^{\rho^\prime \sigma^\prime}\Big] {\cal F}^{[i]}\,,
\eqa
where
\bqa
{\cal F}^{[1]}&=&\frac{3}{2}\big[n^{21}(k)(1 + n^{12}(k+\omega))+n^{12}(k)(1 + n^{21}(k+\omega))\big] \, ,\nonumber \\
{\cal F}^{[2]}&=&n^{31}(k)(1 + n^{13}(k+\omega))+n^{13}(k)(1 + n^{31}(k+\omega))\nonumber \\
&+&\frac{1}{2}\big[n^{12}(k)(1 + n^{21}(k+\omega))+n^{21}(k)(1 + n^{12}(k+\omega))\big]\,.
\eqa
In an intuitive manner, the above result shows that each of the two diagonal gluons in SU(3) leads to a structurally similar contribution to the squared matrix element despite their different screening masses and the associated parton distribution functions. This is actually analogue to their counterparts in Eq.~(\ref{m2offnew}). After contracting the Lorentz indices and including the contributions from the off-diagonal gluons as given by Eq.~(\ref{qhatoffsu3}), our final result for the jet quenching parameter in the presence of a background field is obtained as 
\bqa\label{qhatsu3}
    \hat{q}&=&\frac{g^4}{3(4\pi)^3} \sum_{i=1}^{4}\int_{0}^{\infty} dk \int_{-k}^{\infty} d\omega \int_{|\omega|}^{\min(2k+\omega, \sqrt{\omega^2+\Lambda_{\perp}^2})} dq\,q^2  (1-{\hat \omega}^2) {\cal F}^{[i]}  \nonumber\\
    &\times&\Big[|{\tilde \Delta}_T^{[i]}(Q,{\vec {\sf{q}}})|^2(1-{\hat \omega}^2)^2\left((\omega+2k)^2+3q^2\right)
    +2|{\tilde \Delta}_L^{[i]}(Q,{\vec {\sf{q}}})|^2\left((\omega+2k)^2-2q^2\right)\Big]\,.
\eqa
As we can see, except the complicated color structures in the sums of the distribution functions $ {\cal F}^{[i]}$, the above result can be considered as an extension of the corresponding results at ${\cal Q}=0$. Specifically, we only need to replace the perturbative Debye mass with the ${\cal Q}$-modified ones for each gluons. Such an extension looks very simple, however, according to our above discussions, it is highly nontrivial to show this result, especially for the diagonal part.

Finally, as a useful check, the color factors in Eqs.~(\ref{m2off}) and (\ref{m2dia}) at vanishing background field are found to be $N(N^2-N)/2$ and $N(N-1)/2$, respectively. It shows the correct proportionality between the $N^2-N$ off-diagonal gluons and $N-1$ diagonal gluons.

\subsection{Numerical results and discussions}\label{nre}

\begin{figure*}[t]
    \begin{center}
    \includegraphics[width=0.48\linewidth]{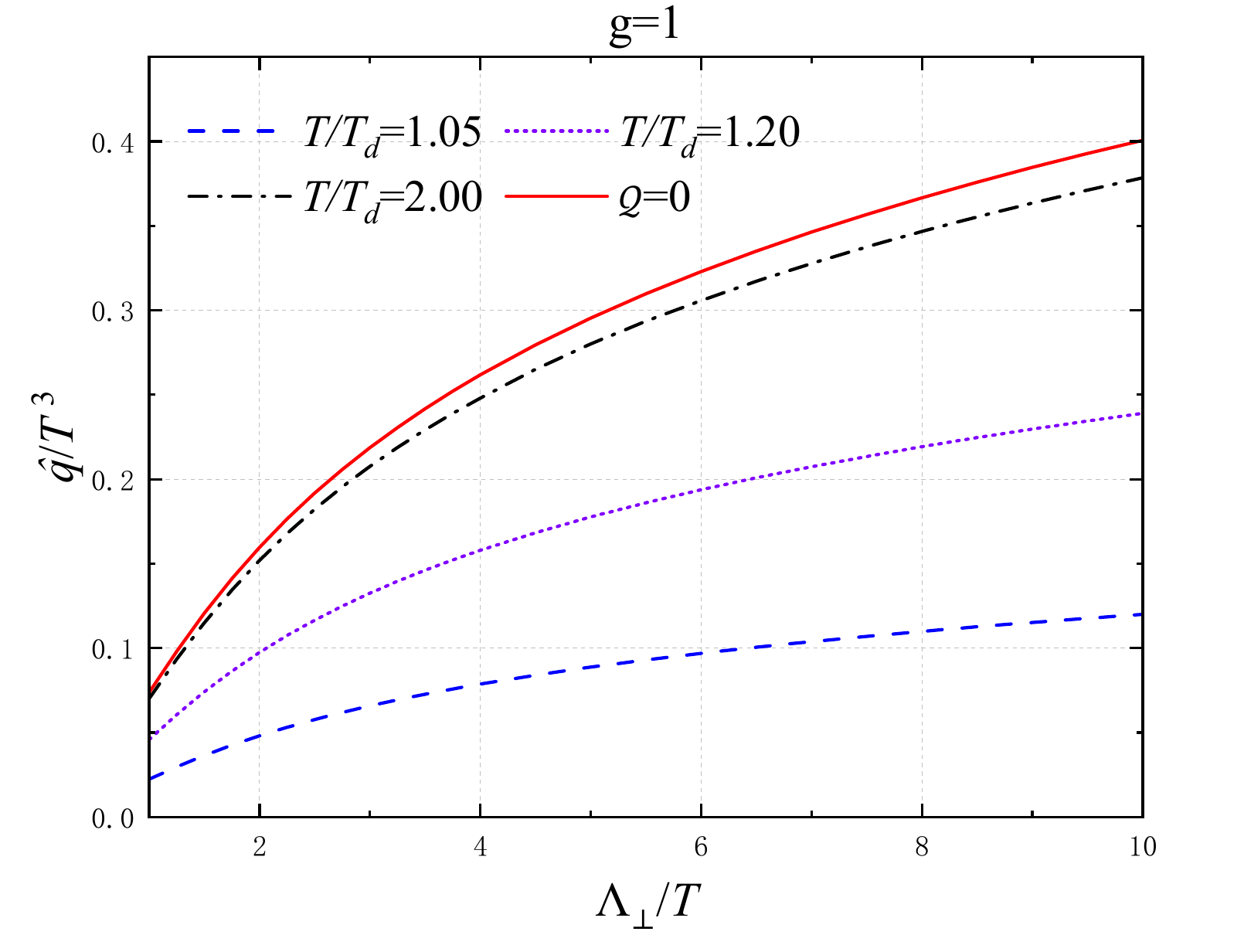}
    \includegraphics[width=0.48\linewidth]{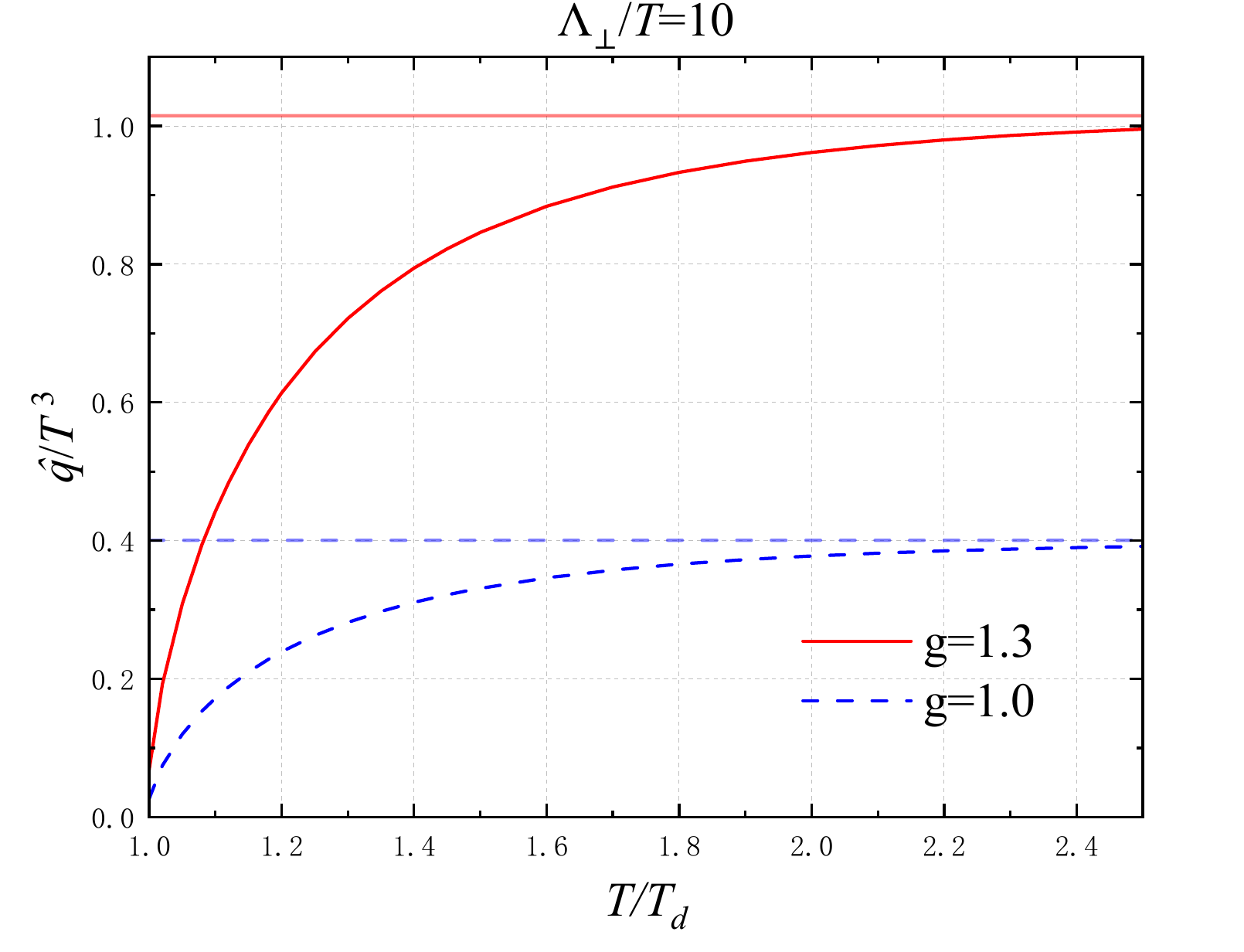}
    \end{center}
    \vspace{-5mm}
    \caption{The dimensionless jet quenching parameter in a background field as a function of the cutoff $\Lambda_\perp/T$ (left) and the temperature $T/T_d$ (right). We use the fixed gauge coupling and also show the corresponding result at ${\cal Q}=0$ for comparison.}
    \label{qhat1}
\end{figure*}

In this subsection, numerical results for the jet quenching parameter based on Eq.~(\ref{qhatsu3}) are presented where the background field is determined according to Eq.~(\ref{bf}). In the left plot of Fig.~\ref{qhat1}, we show the cutoff dependence of the jet quenching parameter rescaled by $T^3$ where a fixed gauge coupling $g=1$ is used. 
As a function of $\Lambda_\perp/T$, the dimensionless ${\hat q}/T^3$ has no $T$-dependence when ${\cal Q}=0$ and the corresponding result is given by the solid curve for reference. After introducing the background field which depends solely on the plasma temperature, we can see a decreased ${\hat q}/T^3$ and the ${\cal Q}$-induced reduction is very significant at low temperature and large cutoff. In general, ${\hat q}/T^3$ increases with the increased cutoff, however, the growth rate declines in the presence of a background field which becomes more visible as the temperature approaches to $T_d$. In the right plot, given the cutoff $\Lambda_\perp/T=10$, we show how the dimensionless jet quenching parameter depends on the temperature. It is found that near the critical temperature, a sharp decrease in ${\hat q}/T^3$ shows up which indicates a strong suppression due to the background field. On the other hand, above $2 \sim 3 T_d$, the influence of the background field is less accentuated and ${\hat q}$ gets very close to its values at vanishing ${\cal Q}$ which is denoted by a horizontal line in this plot. In addition, we consider two different coupling constants and the above findings apply to both cases although the asymptotic values of ${\hat q}$ increase for large gauge couplings.\footnote{Notice that if we plot ${\hat q}/(g^4 T^3)$ instead of ${\hat q}/T^3$, the asymptotic values increase for small gauge couplings.}

\begin{figure*}[t]
    \begin{center}
    \includegraphics[width=0.48\linewidth]{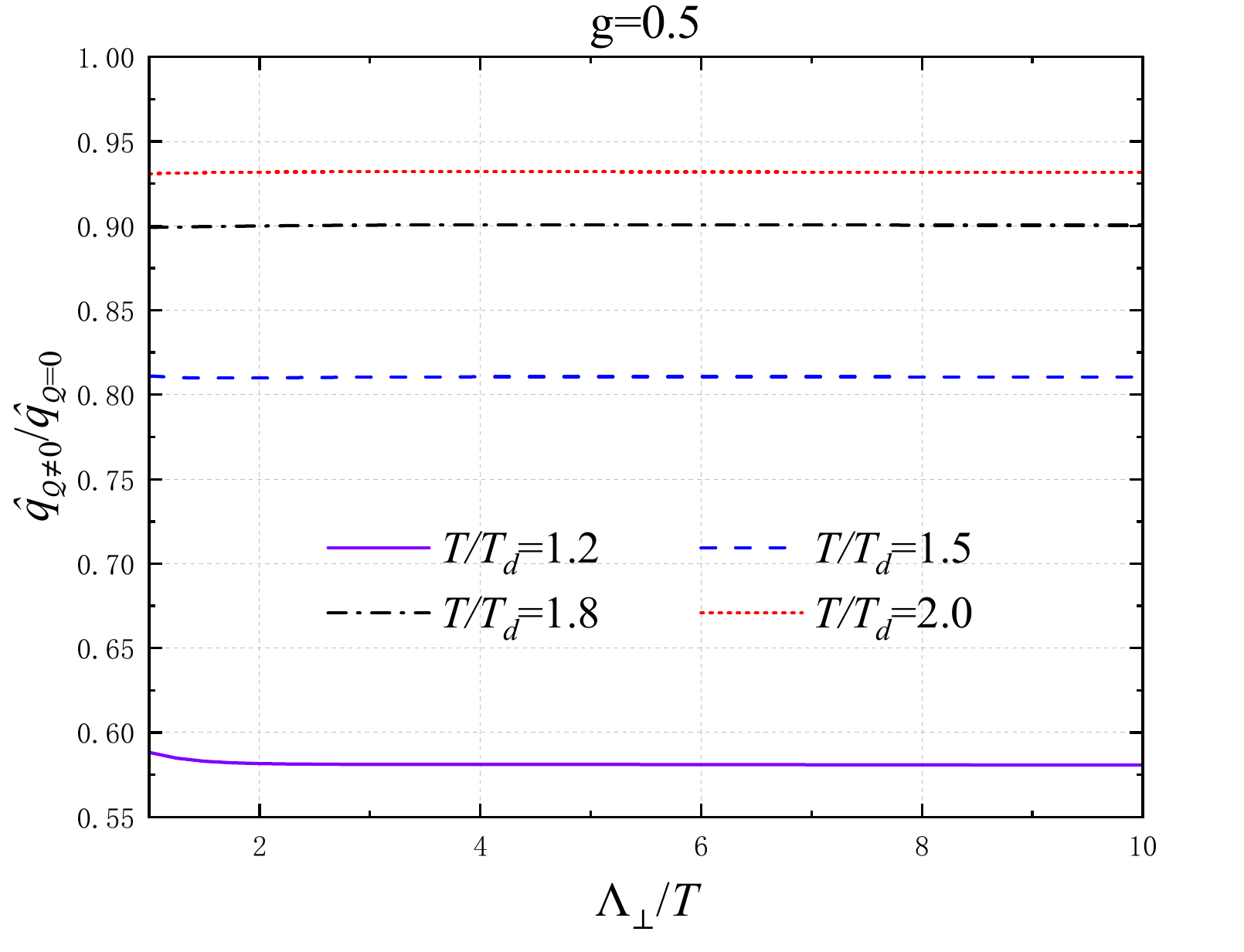}
    \includegraphics[width=0.48\linewidth]{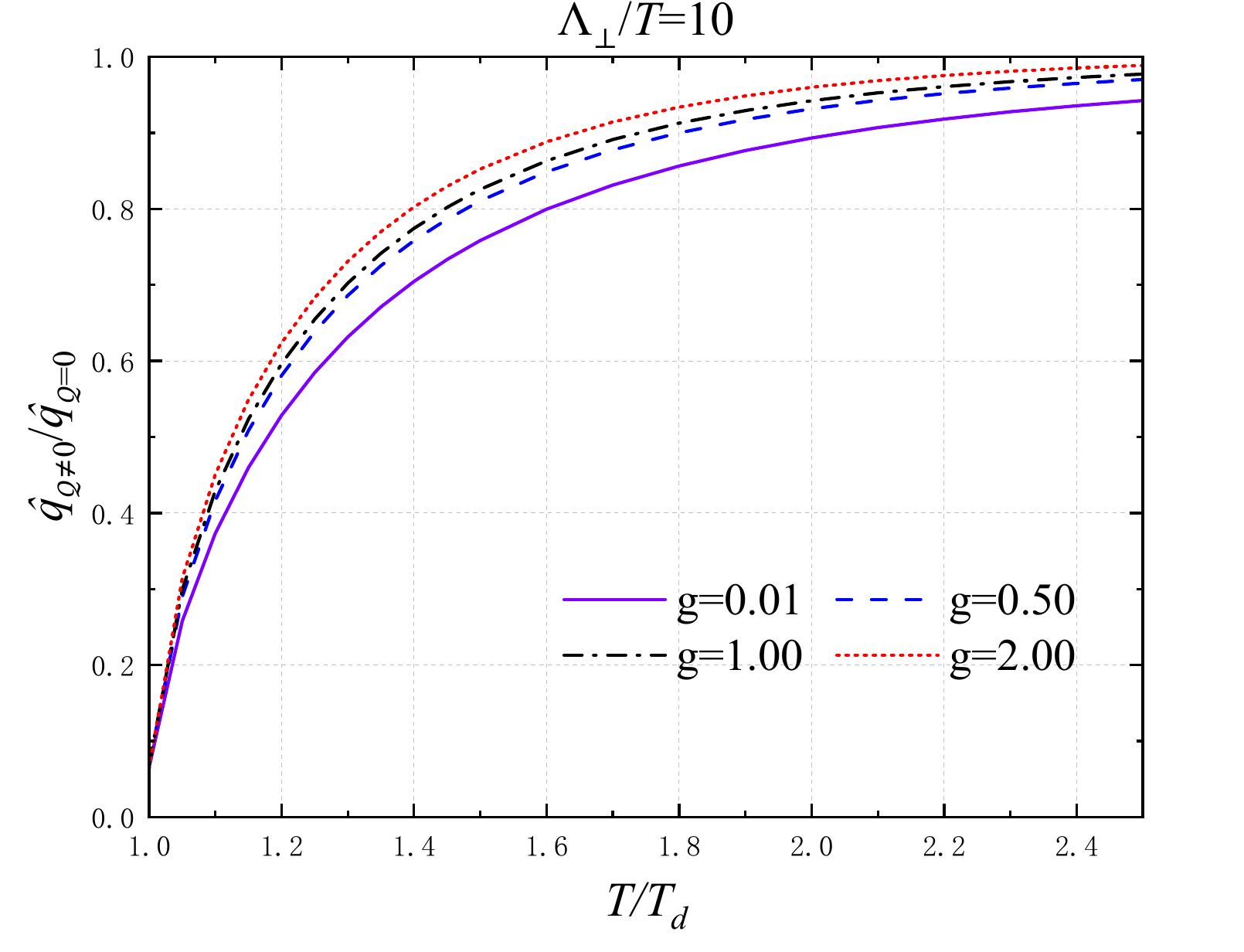}
    \end{center}
    \vspace{-5mm}
    \caption{The ratio of the jet quenching parameter with and without the background field. Left: the cutoff dependence for different temperatures with the gauge coupling fixed. Right:  the temperature dependence for different coupling constants with the cutoff fixed.}
\label{qhatra}
\end{figure*}

A more direct way to illustrate the influence of a nonzero background field is to look at the ratio of the jet quenching parameter with and without $\mathcal{Q}$. The corresponding result can be found in Fig.~\ref{qhatra}. According to the left plot, the ${\hat q}$-ratio does not show a notable change when varying $\Lambda_\perp/T$ although the dimensionless ${\hat q}/T^3$ strongly depends on the cutoff as shown in Fig.~\ref{qhat1}. Therefore, we can neglect the weak dependence on the cutoff so that this ratio depends only on the gauge coupling $g$ and the temperature $T$. In the right plot, choosing $\Lambda_\perp/T=10$, we show the temperature dependence of the ${\hat q}$ ratio for various coupling constants. Roughly speaking, this ratio is almost $1$ above $\sim 2 T_d$ ,which is consistent with our observation in Fig.~\ref{qhat1}. Surprisingly, it becomes only $\sim 5\%$ near the critical temperature and such a dramatic decrease indicates the important role of a nonzero ${\cal Q}$ in determining ${\hat q}$. In addition, this suppression remains significant across the
entire semi-QGP region, from $T_d$ to approximately $2\sim 3 T_d$. In this narrow region, the rapid change of the ${\hat q}$-ratio is actually due to the sharp decrease of ${\cal Q}$ with increasing $T$ which can be obtained based on Eq.~(\ref{bf}). We also consider the temperature dependence of the ${\hat q}$-ratio for different values of $g$, the results exhibit a moderate change when varying gauge coupling. On the other hand, taking into account the fact that the gauge coupling does not change significantly in the semi-QGP,\footnote{Using the two-loop running coupling, we can estimate that $g$ varies from $\sim 1.9$ down to $\sim 1.5$ as the temperature changes from $T_d$ to about $3T_d$.} we can further neglect its dependence on $g$ and take the ${\hat q}$ ratio as a function of the temperature only. As a consequence, the jet quenching parameter in the presence of a background field can be well approximated by multiplying its counterpart at ${\cal Q} = 0$ with an overall temperature-dependent correction factor. 

\begin{figure}[htbp]
\begin{center}        
\includegraphics[width=0.48\linewidth]{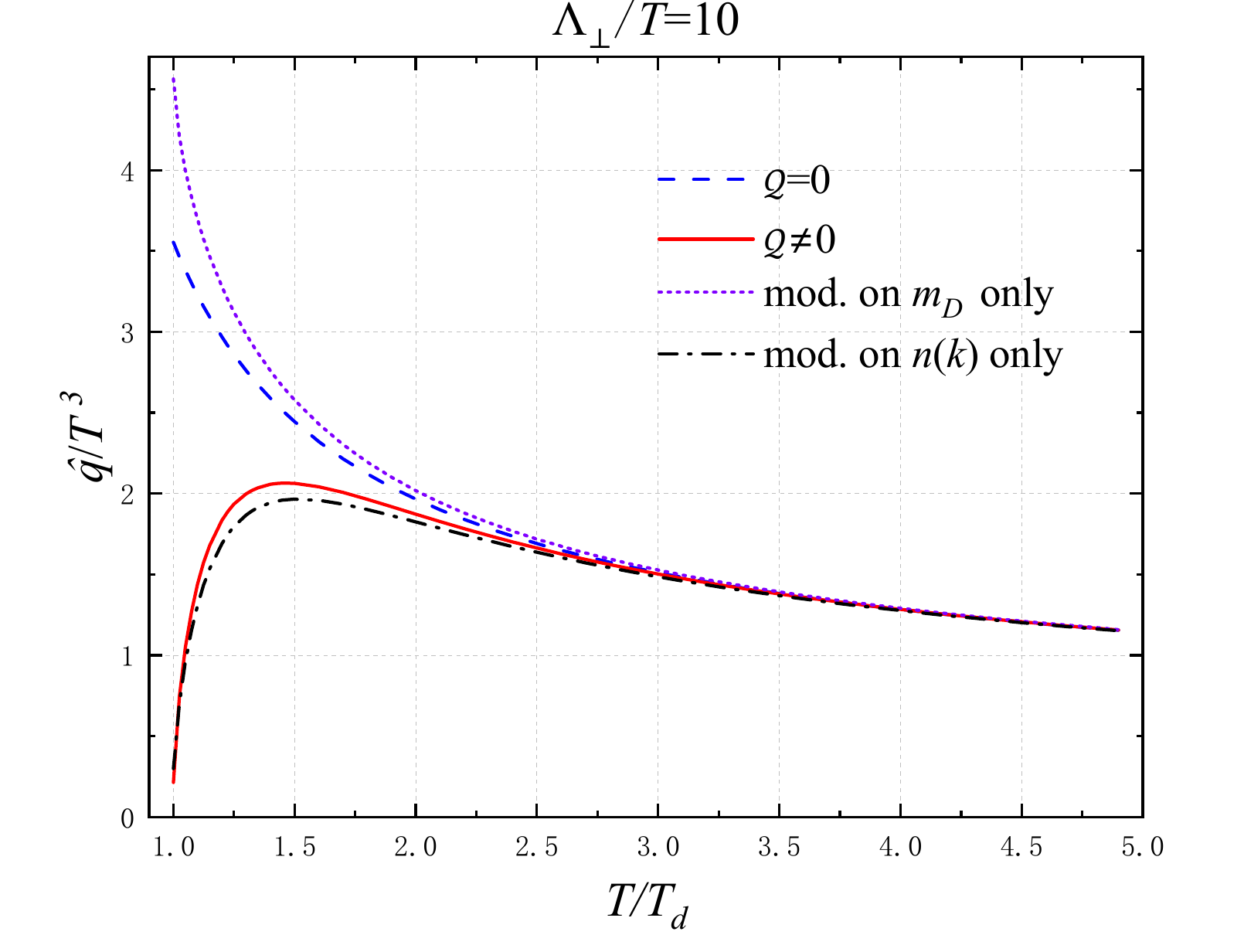}
\includegraphics[width=0.48\linewidth]{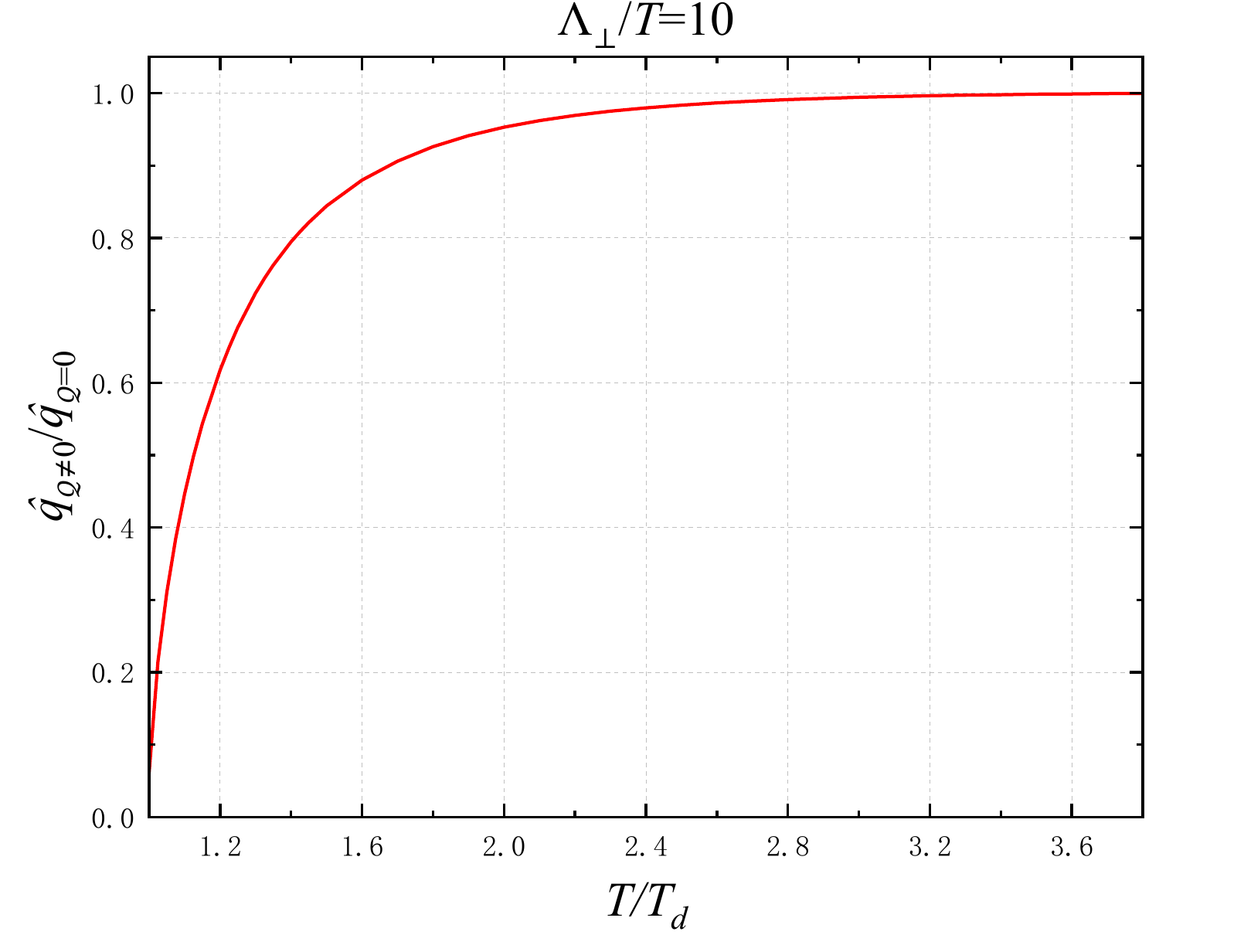}
\end{center}
\vspace{-5mm}
\caption{The jet quenching parameter with running coupling effect. Left: The temperature dependence of the dimensionless ${\hat q}/T^3$ with and without ${\cal Q}$. The partial modification due to the background field is also shown. See text for more details. Right: The temperature dependence of the ${\hat q}$ ratio.}
\label{qhatrun}
\end{figure}

The suppression of the jet quenching parameter due to the nonzero background field could be compensated by the running effect of the gauge coupling which naively enhances ${\hat q}/T^3$ when decreasing the temperature. Therefore, we also evaluate the jet quenching parameter by using the two-loop running coupling which is given by
\beq
g^{-2}(T)=2\beta_0\ln\left(2\pi T/\Lambda_{\overline{MS}}\right)+\frac{\beta_1}{\beta_0}
\ln\left[2\ln\left(2\pi T/\Lambda_{\overline{MS}}\right)\right]\, ,
\label{2loop}
\eeq
with $\beta_0=11/(16\pi^2)$ and $\beta_1=102/(256\pi^4)$. The corresponding results of ${\hat q}$ are presented in Fig.~\ref{qhatrun}. In numerical evaluations, we choose $\Lambda_{\overline{MS}}\approx 350\,{\rm MeV}$. As shown in the left plot, without the background field, ${\hat q}/T^3$ increases as $T$ approaches to $T_d$, this is entirely attributed to the running coupling effect. Including a nonzero background field, we find a nonmonotonic $T$ dependence which can be understood as the following. In the high temperature region where ${\cal Q}$ is small, ${\hat q}/T^3$ grows with decreasing
$T$ because of the increase in the running coupling. Due to the delicate interplay between the running coupling and the background field, there exists a turning point at a temperature just above $T_d$ where ${\hat q}/T^3$ starts to fall. In fact, for temperatures close to $T_d$, influence of a nonzero background field, which suppresses the jet quenching parameter, becomes dominant over the running effect of the gauge coupling which in turn increases the result. We point out that such a $T$-dependence behavior is very similar to that of the ${\cal Q}$-modified screening masses (scaled by the temperature) in the semi-QGP~\cite{Guo:2020jvc}. 

It is also worth noting that the influence of the background field on the jet quenching parameter manifests in two aspects. One is the modification on the distribution function which leads to a reduced ${\hat q}$ if we assume the screening mass remains the perturbative one, i.e., $m_D=g T$; while the other corresponds to the ${\cal Q}$-modified screening mass which, on the other hand, enhances ${\hat q}$ if the ${\cal Q}$ modification on the distribution function has been ignored. These features have been also demonstrated in the left plot of Fig.~\ref{qhatrun}. An intuitive way to understand this behavior is to consider the soft contributions to ${\hat q}$ as given in Eq.~(\ref{qs}). For soft momentum transfer, the general expression Eq.~(\ref{qhat}) is reduced to Eq.~(\ref{qs}) in which the prefactor $m_D^2$ originates from the momentum integral of the distribution function. Therefore, the ${\cal Q}$ modification on the distribution functions (equivalently, on the prefactor $m_D^2$ in this case) causes the decrease of ${\hat q}$. The other $m_D^2$ appearing in the logarithm in Eq.~(\ref{qs}) corresponds to the screening mass in the HTL resummed gluon propagator and the decrease of this mass due to nonzero ${\cal Q}$ in turn leads to an increased ${\hat q}$. In addition, our numerical result suggests that the ${\cal Q}$ modification on the distribution function plays a more important role and the net effect is a reduction in the jet quenching parameter when a nonzero background field is introduced.\footnote{At very high temperatures, the ${\cal Q}$-modification becomes negligible. However, given the cutoff $\Lambda_\perp/T=10$, numerically we find a marginal increase in ${\hat q}$ in the presence of a very small ${\cal Q}$. In this case, increasing the cutoff $\Lambda_\perp/T$ can lead to a reduced ${\hat q}$ again.}

In the right plot of Fig.~\ref{qhatrun}, we present the ratio of the jet quenching parameter under the use of running coupling. The result shows that the ${\hat q}$ ratio is almost unaffected as compared to that obtained by using constant gauge couplings with values around one. For phenomenological use, we can simply introduce a parametrization for the ${\hat q}$ ratio. In terms of the background field ${\sf q}$ in Eq.~(\ref{bf}), the parametrization $1-a\,{\sf q}^2+b\,{\sf q}^3$ with $a\approx 43.89$ and $b\approx 113.74$ can very well reproduce the result shown in this plot. 

To assess the quantitative accuracy of our evaluations on ${\hat q}$, we compare our results with those obtained in previous literature as well as the lattice simulations. Within the perturbation theory~\cite{Qin:2009gw}, ${\hat q}/T^3$ is about $3\sim 5$ for a quark jet with energy $E=10\,{\rm GeV}$ at $T_d=0.27\, {\rm GeV}$, while the corresponding value is found to be slightly smaller based on the effective dynamical quasiparticle model~\cite{Grishmanovskii:2022tpb}. Both are comparable to our result at ${\cal Q}=0$ as shown in Fig.~\ref{qhatrun}, but much larger than the value obtained with ${\cal Q}\neq 0$. At high temperature where $T \sim 1\,{\rm GeV}$, we find ${\hat q}/T^3\approx 1.5$ which is consistent with the results in Refs.~\cite{Qin:2009gw,Grishmanovskii:2022tpb}. In addition, the dimensionless jet quenching parameter extracted from the experimental data lies in a region of $2\sim 5$ for temperatures varying from $0.37$ to $0.54 \,{\rm GeV}$~\cite{JETSCAPE:2021ehl}. Although this result applies to the full QCD case, there is no significant difference as compared to our result in Fig.~\ref{qhatrun}. With a nonzero background field, the most interesting behavior is the strong suppression of ${\hat q}/T^3$ as $T\rightarrow T_d$. This is quite different from other theoretical predictions where an enhancement of ${\hat q}/T^3$ shows up instead. A similar suppression is also found in lattice simulation~\cite{Kumar:2020wvb}. However, as the temperature increases from $T_d$, lattice data shows a much quicker saturation in ${\hat q}/T^3$ as compared to our result based on the background field effective theory, although the asymptotic values at extremely high temperature are very close to each other. We also mention that such a suppression of ${\hat q}/T^3$ near the critical temperature is also observed by using the stochastic vacuum model~\cite{Antonov:2007sh}, but the magnitude of ${\hat q}/T^3$ is much smaller as compared to the lattice data.

\section{Summary}\label{summary}

In this work, we studied the jet quenching parameter by using the background field effective theory. We provided a derivation of ${\hat q}$ with detailed discussions on its highly nontrivial color structure. We also numerically evaluated the jet quenching parameter and analyzed its temperature dependent behavior with focus on the semi-QGP region where dramatic changes were found as compared to the HTL perturbative calculation at vanishing background field. The $T$-dependent background field was determined self-consistently from its equation of motion within the effective theory which characterized the correct temperature dependence of the Polyakov loop. Therefore, our result was expected to give an assessment on how the deconfining phase transition affected the behavior of the jet quenching parameter. 

The theoretical approach adopted to compute the jet quenching parameter involved using the resummed gluon propagator for arbitrary momentum transfer which regulated the infrared divergence and treated the hard and soft processes in a unified framework. It led to a gauge independent ${\hat q}$ that was shown to be consistent to the previous results at vanishing background field. Based on this approach, we derived the jet quenching parameter in the effective theory with the ${\cal Q}$-modified resummed gluon propagator and thermal distribution functions. In the presence of a background field, although the final expression of ${\hat q}$ as given in Eq.~(\ref{qhatsu3}) is formally analogous to its counterpart at ${\cal Q}=0$, the corresponding derivation was not a trivial extension, especially for the contributions from diagonal gluons because of the rather complicated color structures that emerged in the calculation.   

Our numerical results showed that in general a nonzero background field resulted in a reduced ${\hat q}/T^3$. The suppression was significant in the entire semi-QGP, from $T_d$ to $2\sim 3 T_d$. In addition, the ${\cal Q}$ modification on the jet quenching parameter characterized by the ${\hat q}$ ratio was found to be insensitive to the ultraviolet cutoff and exhibited only a moderate dependence on the gauge coupling, therefore, this ratio could be considered as a function of $T$ only.  After taking into account the running coupling effect, the dimensionless ${\hat q}/T^3$ from the background field effective theory showed a nonmonotonic $T$ dependence. In the high temperature region, it increased with decreased $T$ which was attributed to the running coupling. Continuing to lower the temperature, the interplay between the running coupling and the background field gave rise to a turning point at a temperature just above $T_d$ where ${\hat q}/T^3$ sharply fall off due to the influence of the background field. The strong suppression of the dimensionless jet quenching parameter near the critical temperature was also found in the lattice simulation. 

Finally, for phenomenological applications, in order to incorporate the effect of a nonzero background field in the evaluation on ${\hat q}$, it turned out to be feasible to express the jet quenching parameter at ${\cal Q}\neq 0$ by multiplying its counterpart at ${\cal Q}=0$ with a modification factor, which was nothing but the ${\hat q}$ ratio. This ratio could be well described by a parametrization which took a form of a simple polynomial depending only on the background field.

\section*{Acknowledgments}
The work of Y.G. is supported by the NSFC of China under Project No. 12465022; of Q.D. by Guangxi Natural Science Foundation under Grant No. 2023GXNSFBA026027 and by the NSFC of China under Project No. 12305135.

\bibliographystyle{apsrev4-1}
\bibliography{qhatinBF}
\end{document}